\documentclass[aip,jcp,amsmath,preprint,reprint]{revtex4-1}
\usepackage{dcolumn}

\renewcommand{\vec}[1]{{\rm\bf #1}}
\newcommand{\frct}[2]{{\textstyle\frac{#1}{#2}}}
\newcommand{\ka}{\kappa}
\newcommand{\la}{\lambda}
\newcommand{\ep}{\epsilon}
\newcommand{\ps}{\phantom{1}}
\newcommand{\pp}{\phantom{+}}
\newcommand{\de}{{\rm d}}

\begin{document}

\title{Atomic basis functions for molecular electronic structure calculations}

\author{Dimitri N. Laikov}
\email[E-Mail: ]{laikov@rad.chem.msu.ru}
\homepage[Homepage: ]{http://rad.chem.msu.ru/~laikov/}
\affiliation{Chemistry Department, Moscow State University,
119991 Moscow, Russia}

\date{\today}

\begin{abstract}
Electronic structure methods
for accurate calculation of molecular properties
have a high cost that grows steeply with the problem size,
therefore, it is helpful to have
the underlying atomic basis functions
that are less in number but of higher quality.
Following our earlier work [Chem. Phys. Lett. \textbf{416}, 116 (2005)]
where general correlation-consistent basis sets
are defined, for any atom,
as solutions of purely atomic functional minimization problems,
and which are shown to work well for chemical bonding in molecules,
we take a further step here and define
a new kind of atomic polarization functionals,
the minimization of which yields
additional sets of diffuse functions
that help to calculate better molecular electron affinities,
polarizabilities, and intermolecular dispersion interactions.
Analytical representations by generally-contracted Gaussian functions
of up to microhartree numerical accuracy grades
are developed for atoms Hydrogen through Nobelium
within the four-component Dirac-Coulomb theory
and its scalar-relativistic approximation,
and also for Hydrogen through Krypton
in the two-component nonrelativistic case.
The convergence of correlation energy
with the basis set size is studied,
and complete-basis-set extrapolation formulas
are developed.
\end{abstract}

\maketitle

\section{Introduction}

The idea that the molecular electronic structure problem
can be solved in terms of atom-centered basis functions
is as old~\cite{HL27,L29} as the quantum theory of electron itself~\cite{S26,S26b,D28},
but its computational realization has gone a long way
from qualitative pictures to accurate quantitative predictions
of molecular properties and reactivity.
Atomic functions of exponential type~\cite{Z30,S30}
were the first to be used for molecular calculations
at the Hartree-Fock~\cite{H28,F30,R51} (HF)
and limited configuration interaction~\cite{L55} (CI) level;
within the density-functional theory~\cite{KS65},
they are still widely used
thanks to the numerical integration and density-fitting schemes~\cite{BER73}
speeding up the calculations,
and are optimized for all atoms~\cite{LB03}
within the zeroth-order regular approximation~\cite{LBS93}.
Gaussian-type functions are unique
among all classes of elementary functions
in that all multicenter molecular integrals
can be computed analytically~\cite{B50},
and they have become the standard primitive basis
in correlated molecular calculations.
Other kinds of functions can be least-squares fitted
by sums of primitive Gaussians,
this was done first for the exponentials~\cite{HSP69,S70}
and was popular for some time,
but it soon became clear that
energy-optimized Gaussian approximations~\cite{DHP70} of atomic HF wavefunctions
work much better and are a must-have part of a good basis set.
The low variational flexibility of the minimal basis
was understood, so radial~\cite{DHP71,HDP72}
and angular~\cite{HP73} polarization functions
began to be added, often tweaked by hand
to get better computed properties of a favorite set of small molecules.
Later, the so-called diffuse functions~\cite{FP84}
were found to be important, as were multiple polarization functions,
also of higher angular momentum,
and it was around that time
that the need for a more general construction was felt.

The many-body perturbation theory at second~\cite{MP34} (MP2),
third~\cite{B75,PSK77}, fourth~\cite{KP78,KFP80}, and seldom fifth~\cite{RPRH90} order,
as well as the coupled-cluster~\cite{C58,CK60,C66} theory
with single and double~\cite{PB82,HPHRT89} (CCSD)
and perturbative triple~\cite{RTPH89} (CCSD(T)) substitutions
are a family of systematic methods approaching chemical accuracy
in the complete basis limit,
their fifth- to seventh-power scaling with the system size
soon makes the integral evaluation, with its fourth-power scaling,
a small part of the work,
so one can be generous in the choice of the primitive expansion length.
Moreover, there are cubic-scaling integral evaluation techniques
with down to quadratic scaling with the primitive set size:
the pseudospectral decomposition~\cite{F88,RBF90,GRMFLGDR94,TH94,MPF97,IN11}
can even reduce~\cite{MC93}
the scaling of the correlation energy calculation;
the density-fitting (resolution-of-identity) approximation~\cite{SP63,W73,BL77,A88,VAF93}
speeds up the MP2 calculations~\cite{FFK93,WH97},
greatly lowers the memory usage, and is highly parallelizable.

Atomic natural orbitals (ANO) were introduced~\cite{AT87}
for the general contraction~\cite{R73} of Gaussian basis sets,
and this is indeed a general method as, ideally,
only the well-defined atomic solutions would be enough;
for the Hydrogen atom, however,
one had to use the H$_2$ molecule,
and the same would have to be done for all atoms
with one valence electron.
No general way to add the diffuse functions
within the ANO method seems to be found,
the authors resorted~\cite{AT90}
to uncontracting or adding primitive Gaussians
whose exponents may be quite arbitrary.
Unfortunately, the natural occupation numbers have no direct connection
to the correlation energy contribution~\cite{SM17}.

The overwhelming breakthrough in the field
is the development of the now-classic
correlation-consistent basis sets~\cite{D89,WD93} ---
systematic sequences of energy-optimized sets
of single-Gaussian polarization functions
added to generally-contracted minimal HF sets,
with an option for core-core and core-valence~\cite{WD95,PD02} correlation;
the convergence with growing set size
allows the extrapolation~\cite{L65,H85,MT97,HHJKKOW98,V07,HPKW09}
to the (apparent) complete basis set limit;
single-Gaussian diffuse functions optimized
on atomic anions can also be added~\cite{KDH92} ---
even though some atoms have a too small or no electron affinity,
the limited single-Gaussian functional form
helps to get meaningful exponents
for typical molecular applications,
more diffuse even-tempered sets~\cite{WD94}
are used for electrical response properties.
It is remarkable, how well the single-Gaussian form
works for the lighter atoms (up to Ne),
although slightly less so
already for the second row~\cite{DPW01}.
But it is also clear, that a somewhat greater accuracy
can be achieved with the same number of functions
if they are made up from longer primitive sets~\cite{HHT97}.

New approaches are evolving:
a completeness profile~\cite{C95} can be used
to set up a black-box minimization procedure
that yields completeness-optimized~\cite{MV06,LMHH13}
basis sets;
bound virtual states of multiply-ionized atoms
are used as polarization functions
in the numerical grid-based atomic basis sets~\cite{D90}
for density-functional calculations;
polarization-consistent basis sets~\cite{J01,J02b}
are optimized, on a set of prototypical molecules,
for faster convergence of HF energy,
and also~\cite{J02a,VH07} for density-functional calculations;
aiming at specific intermolecular potentials,
interaction-optimized~\cite{DD99} basis sets can be constructed
by direct minimization of the counterpoise-corrected~\cite{BB70}
interaction energy for the dimer.

We have also found~\cite{L05} a new way
to define atomic basis functions
as solutions of atomic variational problems ---
the closed-shell MP2 correlation energy expression
is used as a general minimization functional
to drive the optimization of virtual space.
It is generalized to be applied to all atoms
(even Hydrogen)
by the use of an effective Hamiltonian
constructed from the spherical average-of-configurations
Fock operator,
which can be understood as a model
of an atom in a closed-shell-like molecular environment.
The use of the simplest MP2 functional
seems to be no serious limitation
as can be seen from the comparison~\cite{L05}
with the classic basis sets~\cite{D89,WD93,WD95,PD02}
of the same size in molecular CCSD benchmarks.
(The molecular MP2-optimized virtual space~\cite{AB87}
was shown in tests to be close to optimal
also for the high-level correlated methods
such as CCSD.)
Our method fits naturally to the four-component
wavefunction theories, such as Dirac-Coulomb Hamiltonian
and its scalar-relativistic approximation~\cite{D94},
and yields the atomically-balanced contraction
of kinitically-balanced primitive sets.
All chemically-interesting atoms have been covered
and the sets have been used and worked well
in the studies of compounds
with atoms as heavy as actinides~\cite{SSV07,SS06,UGKMBAU14,UKABLZMGTVU17}.
Still, the limitations show up
when one tries to study intermolecular potentials,
dipole polarizabilities, and negatively-charged
molecular systems ---
the functions optimized for the correlation energy
of the neutral atom are too localized,
and diffuse functions are missing in the set.
We sought a general solution to this problem.
Lowest Rydberg states were tried in this role~\cite{SLBF11},
but are too diffuse for a typical use.
From the known connection between the ionization potential
and the long-range behavior~\cite{HMS69,MPL75}
of the density follows that a minimal set of diffuse functions
(one for each angular symmetry)
cannot fit for all imaginable anionic (or even neutral) species,
and a lower bound for the ionization potential should be set anyway.

After years of thinking,
we found new closed-shell-like polarizability functionals,
derived from a simplified (double) perturbative treatment
of atomic electron affinity at the MP2 level,
whose minimization yields
a first set of diffuse atomic basis functions
that already recover most of the dipole polarizability
and $C_6$ dispersion coefficient and help to get accurate
intermolecular potentials for neutral molecules,
and are also helpful for anionic species
with strong enough electron binding.
We find a way to get further diffuse function sets
to be used for molecules with as low ionization potential
as needed.
All these variational procedures can be used
to contract the primitive sets of Gaussian
(or other well-behaved) functions,
and we have also worked out
a weighted least-squares fitting technique
to optimize the primitive sets of any size
needed to get a given accuracy.
Here we present these new ideas
(after a short review
of our older underlying work~\cite{L05})
and show how they are used
to make a database of atomic basis sets
for all atoms from Hydrogen through Nobelium.
We also study the convergence and extrapolation
towards the complete basis limit
on a few atoms and molecules.

\section{Theory}
\label{sec:theory}

Our general atomic basis functions
can be either four-component spinors of the Dirac-Coulomb theory
(or its scalar-relativistic approximation~\cite{D94}),
or the non-relativistic two-component wavefunctions.
They can be defined
either as solutions of integro-differential equations
or as linear combinations of some primitive functions
whose coefficients are solutions
of constrained functional optimization problems,
the latter algebraic formulation is more practical
and will be given here.

We begin with a set of $N_0$ occupied
wavefunctions $\{\phi_i\}$ that make
the average-of-configurations Hartree-Fock energy
\begin{equation}
\label{eq:E0}
E_0 = \sum\limits_{i=1}^{N_0} w_i H_{ii} + \frct12 \sum\limits_{i,j=1}^{N_0} a_{ij} w_i R_{ii,jj} w_j
\end{equation}
stationary (the orthogonality constraints are always implied).
The one-electron $\{H_{\mu\nu}\}$ and
antisymmetrized two-electron $\{R_{\ka\la,\mu\nu}\}$ integrals
are computed over the given set of $N$ two-component (or $2N$ four-component)
 primitive functions $\{\phi_\mu\}$
and transformed with coefficients $C_{\mu i}$ to those in Eq.~(\ref{eq:E0}).
The occupancies $0 < w_i \leq 1$ and the coupling coefficients $a_{ij} = a_{ji}$
are such as keep the spherical symmetry of a neutral atom,
in the closed-shell case $w_i = 1$, $a_{ij} = 1$.
The stationarity conditions are
\begin{equation}
H_{ui} + \sum\limits_{j=1}^{N_0} a_{ij} R_{ui,jj} w_j = 0
\end{equation}
where $u > N_0$ counts all unoccupied states, and also
\begin{equation}
(w_i - w_j) H_{ij} + \sum\limits_{k=1}^{N_0} (a_{ik} w_i - a_{jk} w_j) R_{ij,kk} w_k = 0.
\end{equation}
The occupied energies
\begin{equation}
\ep_i = F^\textrm{o}_{ii}
\end{equation}
are set to the diagonal elements of the Fock matrix
\begin{equation}
F^\textrm{o}_{ij} = H_{ij} + \frct12 \sum\limits_{k=1}^{N_0} (a_{ik} + a_{jk}) R_{ij,kk} w_k,
\end{equation}
and if the energy of Eq.~(\ref{eq:E0}) is invariant to rotations
for some $ij$-pair, then the condition
\begin{equation}
F^\textrm{o}_{ij} = 0 \qquad\mbox{for}\quad i\neq j,\quad w_i = w_j,\quad a_{ik} w_i = a_{jk} w_j,
\end{equation}
should also be met.
For the virtual subspace, the Fock matrix is taken to be
\begin{equation}
\label{eq:Fv}
F^\textrm{v}_{\mu\nu} = H_{\mu\nu} + \sum\limits_{i=1}^{N_0} R_{\mu\nu,ii} w_i ,
\end{equation}
in the four-component case, it is diagonalized in that subspace
to get $N-N_0$ electronic and $N$ positronic states, the latter being discarded
and all further work is done within the electronic part.

For some atoms, there are low-lying states not included in the average of Eq.~(\ref{eq:E0})
but important for chemical bonding, so $N_1$ functions should be added
to the occupied set $\{\phi_i\}$ to account for it,
and this is done by the diagonalization of a Fock matrix
\begin{equation}
\label{eq:Fp}
F^+_{\mu\nu} = H_{\mu\nu} + \sum\limits_{i=1}^{N_0} R_{\mu\nu,ii} w^+_i
\end{equation}
in the space of $N-N_0$ virtual electronic wavefunctions,
\begin{equation}
\label{eq:ep}
F^+_{ij} = \ep_i \delta_{ij} \qquad\mbox{for}\quad i,j > N_0,
\end{equation}
$N_1$ such states with energies $\ep_i$ go into the occupied set
that from now on has the size $N_\textrm{o} = N_0 + N_1$.
The occupancies $w^+_i$ are for the spherically-averaged configuration
of the singly-ionized atom.

At this time, we build the effective Fock operator
\begin{equation}
\label{eq:F}
\hat{F} =
\sum\limits_{i=1}^{N_\textrm{o}}
 \left|\phi_i\right> \ep_i \left<\phi_i\right|
+ \sum\limits_{i,j=1}^{N_\textrm{o}}
 \bigl(1 - \left|\phi_i\right> \left<\phi_i\right|\bigr)
 \hat{F}^\textrm{v}
 \bigl(1 - \left|\phi_j\right> \left<\phi_j\right|\bigr)
\end{equation}
that is spherically symmetric,
and together with the two-electron interaction
it is all that is needed as input
to a correlation energy calculation.
We \textit{forget} about the fractional occupations ---
now we work with the (pseudo)atom as if it had
the closed-shell configuration
with $N_\textrm{o}$ fully filled levels.

A set of $N_\textrm{v}$ virtual wavefunctions $\{\phi_u\}$
(needed for electron correlation,
atomic polarization upon chemical bonding,
electron affinity, and dispersion interaction in molecules)
can be grown \textit{stepwise} by minimization
of the functionals (shown below) defined
in terms of the linearly transformed set $\{\phi_a\}$
with coefficients $\{C_{ua}\}$ that diagonalize
their block of the Fock matrix
\begin{equation}
\sum\limits_{v=N_\textrm{o}+1}^{N_\textrm{v}} F^\textrm{v}_{uv} C_{va} = \ep_a C_{ua}
\end{equation}
and have energies $\{\ep_a\}$.
The stepwise growth means
that a set of $N_\textrm{v}^{(1)}$ functions $\{\phi_u\}$
is optimized first for one functional;
next, keeping these $N_\textrm{v}^{(1)}$ frozen,
$N_\textrm{v}^{(2)}$ functions are added and optimized
for another functional of all $N_\textrm{v}^{(1)} + N_\textrm{v}^{(2)}$
functions, and so on.
(Sometimes (see Section~\ref{sec:calc}) the very first set
of $N_\textrm{v}^{(0)}$ functions is taken as the next lowest energy
states of Eq.~(\ref{eq:ep}) after the lowest $N_1$.)

The minimization of a model second-order correlation energy functional
\begin{equation}
\label{eq:E2}
E_2 = \frct14 \sum\limits_{ijab}
 \frac{p_{ij} \left|R_{ai,bj}\right|^2}{\ep_i + \ep_j - \ep_a - \ep_b}
\end{equation}
with respect to some members of the set $\{\phi_u\}$
yields functions nearly optimal
for the most part of electron correlation in atoms and molecules.
The factors $p_{ij}$ in Eq.~(\ref{eq:E2}) are set to 1 or 0
to switch the valence-only, core-valence, and core-core correlation.
The stationarity conditions can be derived
by chain-rule differentiation
with respect to rotations that mix the external $\{\phi_x\}$
and the virtual $\{\phi_u\}$ functions, and can be written as
\begin{equation}
\label{eq:Gxu}
G_{xu} =
\sum\limits_a \left(G_{xa} + \sum\limits_b F_{xb} D_{ab} \right) C_{ua} = 0
\end{equation}
with
\begin{equation}
G_{xa} = \sum\limits_{ijb} \frac{p_{ij} R_{xi,bj} R_{ai,bj}}{\ep_i + \ep_j - \ep_a - \ep_b},
\end{equation}
\begin{equation}
D_{ab} = \sum\limits_{ijc}
\frac{p_{ij} R_{ai,cj} R_{bi,cj}}
{(\ep_i + \ep_j - \ep_a - \ep_c)(\ep_i + \ep_j - \ep_b - \ep_c)}.
\end{equation}
The second derivatives of Eq.~(\ref{eq:E2}) can also be derived
and used in an efficient quadratically-convergent Newton-Raphson optimization
with a careful steps size control.

Until now, we have followed our earlier work~\cite{L05},
but since then the experience has shown that
such atomic basis sets lack \textit{diffuse} functions 
needed for accurate calculation of electron affinities,
polarizabilities, and dispersion interactions in molecules.
Now, we have found a model polarization functional
\begin{equation}
\label{eq:Ea}
E_\textrm{a} = \sum\limits_{ia} \frac{p_i \left|\bar{U}_{ai} \right|^2}{\ep_i - \ep_a},
\end{equation}
\begin{equation}
\label{eq:U}
\bar{U}_{\mu\nu} = \sum\limits_{i=1}^{N_0} R_{\mu\nu,ii} \bar{w}_i,
\end{equation}
whose minimization yields a set of functions, with angular momenta
up to those in the occupied set, that help to account for the changes
upon electron attachment (or detachment).
The occupancies $\bar{w}_i$ add up to one
and are spherically-averaged,
typically $\bar{w}_i = w_i - w^+_i$,
the switching factors $p_i$ are set to 1 or 0.
Eq.~(\ref{eq:Ea}) can be understood as the second-order perturbative
correction for the averaged electron attachment energy
at the Hartree-Fock level,
and we believe it to be a better functional
for driving the diffuse function optimization
than the directly computed energy of an atomic anion ---
some atoms have a tiny or no electron affinity
so the functions may become too diffuse,
but the perturbative first order changes are
more localized and still in the right direction.
The stationarity conditions are as in Eq.~(\ref{eq:Gxu})
but now with
\begin{equation}
G_{xa} = \sum\limits_i \frac{p_i \bar{U}_{xi} \bar{U}_{ai}}{\ep_i - \ep_a},
\end{equation}
\begin{equation}
D_{ab} = \sum\limits_i \frac{p_i \bar{U}_{ai} \bar{U}_{bi}}{(\ep_i - \ep_a)(\ep_i - \ep_b)},
\end{equation}
and there is a simple explicit solution.

To optimize the diffuse functions with higher-than-occupied angular momenta,
we model the changes in the MP2 correlation energy upon electron attachment.
The perturbed occupied wavefunctions
\begin{equation}
\label{eq:fa}
\bar{\phi}_i = \phi_i + \sum\limits_{a=N_\textrm{o} +1}^N \phi_a \frac{\bar{U}_{ai}}{\ep_i - \ep_a},
\end{equation}
with the sum running here
over the whole $N-N_\textrm{o}$ virtual space in diagonal representation
of the matrix of Eq.~(\ref{eq:Fv}), are computed first,
and then the set $\{\bar{\phi}_i\}$ is used in Eq.~(\ref{eq:E2}) instead of $\{\phi_i\}$
to get the new functional $\bar{E}_2$. The difference
\begin{equation}
\label{eq:DE2}
\Delta \bar{E}_2 = \bar{E}_2 - E_2
\end{equation}
is a measure of how the members of the set $\{\phi_u\}$ under optimization
help to lower the correlation energy of the atomic anion
more than of the neutral atom,
and this is the functional that is minimized to get
the set of diffuse functions.
With the intermediate normalization of Eq.~(\ref{eq:fa})
one term in $\bar{E}_2$
is exactly canceled by $E_2$, and we like it.

Here we can stop, as we now have enough tools
to build systematic sequences of atomic basis set
for a broad range of molecular applications.
Still, we should be aware of their limitations
and would like to take a look ahead
towards a better sampling
of the diffuse tail region.
We have studied the dipole polarizability functional
\begin{equation}
\label{eq:Ed}
E_\textrm{d} = \sum\limits_{ia} \frac{p_i \left|\vec{r}_{ai} \right|^2}{\ep_i - \ep_a},
\end{equation}
and the homoatomic $C_6$ dispersion coefficient functional
\begin{equation}
\label{eq:E6}
E_6 = \frct23 \sum\limits_{ijab}
 \frac{p_{ij} \left|\vec{r}_{ai} \right|^2 \left|\vec{r}_{bj} \right|^2}
      {\ep_i - \ep_j - \ep_a - \ep_b},
\end{equation}
where $\vec{r}_{ai}$ are the dipole moment integrals,
minimization of either of them yields a set of functions
that are somewhat more diffuse
than those based on Eqs.~(\ref{eq:Ea}) and~(\ref{eq:DE2}) ---
these can be added to the set after the former ones,
but our molecular tests show this to be of little help
for most molecular systems,
even for noble gas dimers there is only a small
lowering of the potential energy curve.
Higher multipole analogs of Eqs.~(\ref{eq:Ed}) and~(\ref{eq:E6})
could have been studied,
but we do not feel it to be the right way forward.

A smooth sampling of the diffuse tails
can be done
with a one-parameter family of model Fock operators
\begin{equation}
\hat{F}_\zeta = \hat{F} + \zeta \hat{\bar{U}},
\end{equation}
with $\hat{\bar{U}}$ from Eq~(\ref{eq:U})
and $0\le \zeta \le \zeta_\textrm{max}$,
that can be diagonalized (with a frozen core option)
to get the wavefunctions of the form
\begin{equation}
\label{eq:fz}
\bar{\phi}_{i\zeta} =
\phi_i + \sum\limits_{a=N_\textrm{o} +1}^N \phi_a \; \bar{C}_{ai\zeta}
\end{equation}
with energies $\bar{\ep}_{i\zeta}$,
and $\zeta_\textrm{max}$ can be found such
that $\bar{\ep}_{N_\textrm{o} \zeta_\textrm{max}} = \ep_\textrm{max}$.
A good $\ep_\textrm{max} < 0$ can be set, for all atoms,
to grow the tails as diffuse as $\exp(-\sqrt{-2\ep_\textrm{max}}\; r)$
for the new members of the set $\{\phi_u\}$,
such that if the $\hat{F}_\zeta$ is diagonalized
in the subspace to get the wavefunctions
\begin{equation}
\label{eq:fzu}
\tilde{\phi}_{i\zeta} =
\phi_i + \sum\limits_{u=N_\textrm{o}+1}^{N_\textrm{v}} \phi_u \; \tilde{C}_{ui\zeta}
\end{equation}
with energies $\tilde{\ep}_{i\zeta}$,
then the integral
\begin{equation}
\label{eq:Ez}
\tilde{\ep} =
\int\limits_0^{\zeta_\textrm{max}}
 \sum\limits_i \tilde{\ep}_{i\zeta}
 \de \zeta
\end{equation}
is minimized. It can be seen that the functional of Eq.~(\ref{eq:Ea})
is a lowest-order perturbative approximation
to that of Eq.~(\ref{eq:Ez}),
and the functions of Eq.~(\ref{eq:fa}) are nothing else
than first-order perturbative analogs of those of Eq.~(\ref{eq:fz}).
In the same way,
to get the higher-than-occupied functions,
the integral
\begin{equation}
\Delta \bar{E}_2 =
\int\limits_0^{\zeta_\textrm{max}}
 \left( \bar{E}_2(\zeta) - E_2 \right)
 \de \zeta
\end{equation}
can be minimized,
where $\bar{E}_2(\zeta)$ is built
upon the set $\{\bar{\phi}_{i\zeta}\}$
instead of $\{\phi_i\}$ in $E_2$ of Eq.~(\ref{eq:E2}).

The general atomic basis functions outlined above
can be computed to any meaningful accuracy
by numerically solving the underlying variational problems.
Once the (nearly) exact solutions $\{\phi_k\}$
are at hand, practical approximations,
such as the traditional contracted Gaussian or exponential (Slater-type)
functions, can be developed for use in molecular calculations.
First, we optimize the exponents of the primitive functions
by weighted least-squares fitting, minimizing
\begin{equation}
\label{eq:q}
Q =
\sum\limits_{k=1}^{N_\textrm{o} + N_\textrm{v}}
 \beta_k \int \left|\tilde{\phi}_k (\vec{r}) - \phi_k (\vec{r}) \right|^2
 w(\vec{r})\; \de^3 \vec{r}
\end{equation}
with respect to both
the linear coefficients $\{\tilde{c}_{nk}\}$
and the primitive exponents $\{\alpha_n\}$
of the approximate functions $\{\tilde{\phi}_k\}$,
with the weights
\begin{equation}
\beta_k = \left\{
\begin{array}{ll}
\beta_\textrm{o}, &
 k \le N_\textrm{o} \\
1, & k > N_\textrm{o}
\end{array}
\right.
\end{equation}
made heavier for the occupied set
by $\beta_\textrm{o}$,
and the radial weight functions
$w(\vec{r}) = 1/|\vec{r}|$.
After much experimentation,
we set $\beta_\textrm{o}=2^{12}$
that gives a good balance
between the HF and correlation energies,
and our choice of $w(\vec{r})$
leads to the exponents that are close
to the energy-optimized values in the HF case.
We have tried to put the orthonormality constraints
on $\{\tilde{\phi}_k\}$ in the fitting,
but found it to heavily complicate and slow down
the computations without giving better results,
so we put it aside.

There is a known pitfall in the work with
the exponents --- some of them may be driven
towards the same value whereas the linear
coefficients of opposite sign
go towards infinity --- and this is the true
but numerically unstable and impractical solution.
To overcome this, we put a bound
on the closeness and parametrize the exponents as
\begin{equation}
\alpha_n = \exp\left(p_1 + \sum\limits_{m=2}^n \sqrt{p_0^2 + p_m^2} \right),
\end{equation}
and $\{p_n\}$, $n \ge 1$, are now taken
as the optimization variables instead of $\{\alpha_n\}$,
and thus $\alpha_{n+1}/\alpha_n \ge \exp(p_0)$,
we settle on $p_0 = (\ln 2)/4$ as a good compromise
between accuracy and stability.

After the primitive exponents have been optimized,
we run a variational calculation to get the linear coefficients,
and so we get our atomic basis functions
with the least-squares fitted exponents
and energy-optimized linear coefficients.

\section{Calculations}
\label{sec:calc}

We have written a computer code for solving
the atomic variational problems of section~\ref{sec:theory}
with full use of spherical symmetry ---
the angular degrees of freedom are integrated out analytically
and only the radial equations are worked with.
Extended precision floating-point arithmetics
with 256- or 128-bit mantissa is implemented
using the \texttt{X86\_64} 64-bit integer instruction set
(with the wide multiply)
in our hand-written assembly code for high speed ---
this overcomes the severe round-off errors
arising from the near-linear dependence
in a large primitive basis set of densely-spaced
Gaussian functions, and the final results can be reliably
rounded to the standard 64-bit precision.

We use the newest estimate~\cite{AHKN12,HFG11} of the speed of light
$c=137.035999173$
in all relativistic four-component calculations,
and also the finite nucleus model~\cite{VD97}
with Gaussian charge distribution with exponent (in au)
\begin{equation}
\alpha = \frct32 \left(\frac{529177249}{5700 + 8360\cdot\sqrt[3]{M}}\right)^2,
\end{equation}
where $M$ is the (integer) mass number of the most abundant isotope;
point nucleus is used in the non-relativistic case.
We solve the variational problems of Section~\ref{sec:theory}
to a very high accuracy over a huge even-tempered
primitive Gaussian basis with exponents
\begin{equation}
\alpha_p = 2^{p/3}
\end{equation}
where $-69\le p \le 111$,
that makes 181 radial functions for each angular symmetry;
we estimate the overlap between these and the exact solutions
to be of the order $1 - 2^{-64}$ for the occupied set
and somewhat less for the virtual.
In the non-relativistic case, a range $-69\le p \le 225$
would be needed to meet the nuclear cusp condition,
but we find it more practical to use
\begin{equation}
\alpha_p = 2^{p/3} + \exp(ap - b)
\end{equation}
with $a$ and $b$ optimized for each atom
on its hydrogen-like ion with one electron,
and a more narrow range of $p$.

The standardized electronic configurations of atoms
are shown in Table~\ref{tab:conf},
where $L^+$ is the angular momentum for which one
electron is removed to get
the occupancies $w^+_i$ in Eq.~(\ref{eq:Fp}),
$L_1$ is for the $N_1$ functions of Eq.~(\ref{eq:ep})
added to the occupied set,
$L_2$ is for the first $N_\textrm{v}^{(0)}$ virtual
functions also from Eq.~(\ref{eq:ep}),
and the atoms are marked for which the diffuse functions
may be added.
To get the coupling coefficients in Eq.~(\ref{eq:E0}),
we use the average level~\cite{GMP76} formalism
in the Dirac-Coulomb case,
and also the usual average of the highest-spin configurations
in the scalar-relativistic and non-relativistic cases.

The stepwise optimization begins with the outermost
occupied shell block (of the same principal quantum number $n$),
for which the functional of Eq.~(\ref{eq:E2}) with valence-only $p_{ij}$
is minimized first
to get a set of virtual functions with $N^\textrm{v}_{ln}$ radial parts
for each angular momentum $l$,
then a set of diffuse functions may be optimized
using Eq.~(\ref{eq:Ea}) followed by Eq.~(\ref{eq:DE2});
the next block of inner valence (for transition metals)
or core shells may then be taken and the virtuals
optimized using Eq.~(\ref{eq:E2})
with $p_{ij}$ set to core-core and core-valence correlation,
and so on to the innermost core shells.
The correlation-consistent virtual set sizes
at each step are taken to be
\begin{equation}
\label{eq:N}
N^\textrm{v}_{ln} = \max\bigl(\la_n - \max(l - l_n^\textrm{max} - 1, 0), 0 \bigr)
\end{equation}
where $l_n^\textrm{max}$ is the highest occupied angular momentum
of the $n$-block,
and $\la_n$ is the set number ($1,2,\dots$) for this $n$ ---
thus $l$ reaches up to $l_n^\textrm{max} + \la_n$
and there are $\la_n$ radial parts for $l \le l_n^\textrm{max} + 1$.
The size of the diffuse set is simply $N^\textrm{v}_{ln} = 1$
for $l \le l_n^\textrm{max} + \la_n$.
The same $\la_n$ is typically used for all $n$,
but our code can work with any other settings.

For most metal atoms, the outermost core shells should be unfrozen,
Table~\ref{tab:conf} shows the number of the $n$-blocks
that are always correlated.
One may also note the atoms Ca, Sr, Ba, and Ra
to have a shell added ($L_2$) to the unoccupied set,
we found it to be the key to get the accurate bonding
properties of these ``subtransition'' metals,
to cure the known pathology~\cite{WVKBS00}.

\begingroup
\squeezetable
\begin{table}[h]
\caption{\label{tab:conf}Atomic electronic configurations.}
\begin{ruledtabular}
\begin{tabular}{lllllll}
atoms    & number of electrons$^a$ & $L^+$ & $L_1$ & $L_2$ & diffuse$^b$ & $n$$^c$ \\
\hline
H, He    & $\ps 1+                        $ & 0 &   &   & + & 1 \\
Li, Be   & $\ps 3+                        $ & 0 & 1 &   &   & 2 \\
B -- Ne  & $\ps 4\pp,\ps 1+               $ & 1 &   &   & + & 1 \\
Na, Mg   & $\ps 5+  ,\ps 6                $ & 0 & 1 &   &   & 2 \\
Al -- Ar & $\ps 6\pp,\ps 7+               $ & 1 &   &   & + & 1 \\
K        & $\ps 7\pp,   12                $ & 0 & 1 &   &   & 2 \\
Ca       & $\ps 8\pp,   12                $ & 0 & 1 & 2 &   & 2 \\
Sc -- Zn & $\ps 8\pp,   12\pp,\ps 1+      $ & 0 & 1 &   &   & 2 \\
Ga -- Kr & $\ps 8\pp,   13+  ,   10       $ & 1 &   &   & + & 1 \\
Rb       & $\ps 9\pp,   18\pp,10          $ & 0 & 1 &   &   & 2 \\
Sr       & $   10\pp,   18\pp,10          $ & 0 & 1 & 2 &   & 2 \\
Y -- Cd  & $   10\pp,   18\pp,11+         $ & 0 & 1 &   &   & 2 \\
In -- Xe & $   10\pp,   19+  ,20          $ & 1 &   &   & + & 1 \\
Cs       & $   11\pp,   24\pp,20          $ & 0 & 1 &   &   & 2 \\
Ba       & $   12\pp,   24\pp,20          $ & 0 & 1 & 2 &   & 2 \\
La       & $   12\pp,   24\pp,21          $ & 0 & 1 &   &   & 2 \\
Ce -- Yb & $   12\pp,   24\pp,21\pp,\ps 1+$ & 0 & 1 &   &   & 3 \\
Lu -- Hg & $   12\pp,   24\pp,21+  ,   14 $ & 0 & 1 &   &   & 2 \\
Tl -- Rn & $   12\pp,   25+  ,30\pp,   14 $ & 1 &   &   &   & 1 \\
Fr       & $   13\pp,   30\pp,30\pp,   14 $ & 0 & 1 &   &   & 2 \\
Ra       & $   14\pp,   30\pp,30\pp,   14 $ & 0 & 1 & 2 &   & 2 \\
Ac       & $   14\pp,   30\pp,31\pp,   14 $ & 0 & 1 &   &   & 2 \\
Th -- No & $   14\pp,   30\pp,31\pp,   15+$ & 0 & 1 &   &   & 3 
\end{tabular}
\begin{flushleft}
$^a$ Number of electrons for each angular momentum;
the plus $+$ means the number grows in the row
starting with the given value. \\
$^b$ Whether the diffuse functions are added. \\
$^c$ At least $n$ of the outermost
principal quantum number shell blocks
are correlated.
\end{flushleft}
\end{ruledtabular}
\end{table}
\endgroup

With the nearly-exact solutions at hand,
the least-squares optimization based on Eq.~(\ref{eq:q})
is run to get the primitive basis sets
of growing size --- this can be done either
separately and independently for each angular momentum $l$,
or all at once with the exponents
shared between all $l$.
The former is more flexible, economical, and natural,
so we do most of our work this way;
but the latter is helpful to speed up
some electronic structure methods
if the integral evaluation
can make use of shared exponents,
so we also do it on a smaller scale.
The non-linear optimization of exponents
sometimes finds multiple minima,
and we often had to feed it
with several sets of starting values,
made by hand, to get either the lowest error
or, more seldom, a more regular variation
of the exponents with the atomic number.
We have always seen the close-to-exponential convergence
of the fit error with the primitive set size $M_l$,
and the same is also true
for both the HF- and MP2-energy errors,
which we estimate for a given $M_l$
by running atomic calculations
on the sets of size
$\{\bar{M}_0,\dots,\bar{M}_{l-1},M_l,\bar{M}_{l+1},\dots,\bar{M}_{l_\textrm{max}}\}$
and
$\{\bar{M}_0,\dots,\bar{M}_{l_\textrm{max}}\}$
and subtracting the energies,
where $\bar{M}_l$ are big enough.
Thus we \textit{grade} the primitive sets
by their energy errors $\mathcal{E}_l(M)$ for each $l$,
and now we have to decide
on the standard set sizes $\{M_l^{(\ka)}\}$
so that
$\mathcal{E}_l\left(M_l^{(\ka)}\right)\approx \mathcal{E}^{(\ka)}$
for all $l$ and for both HF and MP2,
that is, the errors are of the same order,
and for $\ka=1,2,\dots$ we should have
the errors close to a geometric series
$\mathcal{E}^{(\ka)}/\mathcal{E}^{(\ka-1)}\approx \varepsilon$.
As a guide, we look at the MP2 values of
$\mathcal{E}_{l_\textrm{max}}(\ka)$ for $\ka=1,\dots,5$
for atoms He and Ne, and we see
$\varepsilon\approx \frct18$,
and so we align all our sets for all atoms.

For $\la_n = 1,2,3,4$ (and $5$ for lighter atoms) in Eq.~(\ref{eq:N})
and $\ka = 1,2,3,4,5$, with and without the correlation
of the outermost or all core electrons,
we have optimized the series of basis sets
for atoms H through No (7682 in number)
for both the Dirac-Coulomb Hamiltonian
and its scalar-relativistic approximation,
and also for H through Kr for the non-relativistic case.
It took us years of hard work at which we grew old and sick,
so we cannot give all details here.
Instead, the files in the supplementary material~\cite{SM}
hold all our data sets (to 64-bit precision)
and everyone is welcome to use them
or to study their properties.

Our mnemonics for the optimized atomic basis sets are:
\begin{itemize}
\item
\texttt{L}$\la$\texttt{\_}$\ka$
for the valence-only correlation,
\item
\texttt{L}$\la$\texttt{a\_}$\ka$
the same with the diffuse functions,
\item
\texttt{L}$\la\la$\texttt{\_}$\ka$,
\texttt{L}$\la\la\la$\texttt{\_}$\ka$
with the outermost core shells
included into the correlation,
\item
\texttt{L}$\la\la$\texttt{a\_}$\ka$,
\texttt{L}$\la\la\la$\texttt{a\_}$\ka$
the same with the diffuse functions,
\item
\texttt{Lx}$\la$\texttt{\_}$\ka$
for all-electron correlation,
\item
\texttt{Lx}$\la$\texttt{a\_}$\ka$
the same with the diffuse functions.
\end{itemize}
It would have been good to test all these
on a set of molecules,
but here we will only study the convergence
and extrapolation towards the complete basis limit
on a few simple but characteristic examples.

\begingroup
\squeezetable
\begin{table}[h]
\caption{\label{tab:e2at}MP2 correlation energy of closed-shell atoms.}
\begin{ruledtabular}
\begin{tabular}{rlllll}
      & \multicolumn{2}{c}{He} & \multicolumn{2}{c}{Ne} & \multicolumn{1}{c}{Ni} \\
$\la$ & \multicolumn{1}{c}{\texttt{L}$\la$}
      & \multicolumn{1}{c}{\texttt{L}$\la$\texttt{a}}
      & \multicolumn{1}{c}{\texttt{L}$\la$}
      & \multicolumn{1}{c}{\texttt{L}$\la$\texttt{a}}
      & \multicolumn{1}{c}{\texttt{L}$\la$} \\
\hline
 1& -0.02887747 & -0.03039747 & -0.2175501 & -0.2354309 & -0.789926 \\
 2& -0.03418358 & -0.03473452 & -0.2751513 & -0.2826869 & -0.998901 \\
 3& -0.03585425 & -0.03610341 & -0.2986350 & -0.3019700 & -1.102947 \\
 4& -0.03653868 & -0.03666732 & -0.3081867 & -0.3099155 & -1.147565 \\
 5& -0.03686850 & -0.03694147 & -0.3128597 & -0.3138398 & -1.169400 \\
 6& -0.03704621 & -0.03709066 & -0.3153930 & -0.3159914 & -1.181063 \\
 7& -0.03715014 & -0.03717877 & -0.3168867 & -0.3172725 & -1.187980 \\
 8& -0.03721486 & -0.03723414 & -0.3178226 & -0.3180828 & -1.192285 \\
 9& -0.03725721 & -0.03727069 & -0.3184385 & -0.3186205 & -1.195121 \\
$\dots$ \\
$\infty$&
    -0.0373774  &             & -0.32021   &            & -1.2032   
\end{tabular}
\end{ruledtabular}
\end{table}
\endgroup

The atoms He, Ne, and Ni in Table~\ref{tab:e2at}
are archetypal for the closed-shell correlation,
and we took pains to go up to $\la = 9$
to come close to the asymptotic behavior.
(For Ni, the sets are not for the standard configuration
of Table~\ref{tab:conf} but for the closed-shell state
with 18 outer electrons correlated.)

A natural functional form
\begin{equation}
\label{eq:ean}
E_\la \approx E_\infty + \sum\limits_{p=3}^{P} \frac{A_p}{(\la + \nu)^p}
\end{equation}
with some small $P>3$
can be used to fit the computed $E_\la$ for a range of $\la$
and thus to get an estimate of $E_\infty$.
The only nonlinear parameter $\nu$ in Eq.~(\ref{eq:ean})
can be adjusted each time,
but we find $\nu = \frct32$ to be a good fixed value,
often very close to the optimal one,
and we set it so everywhere in the following.
By fitting through $P-1$ points $\la = \la_0,\dots,\la_0+P-2$,
each time for a higher $\la_0$,
we can also estimate the residual error of the last $E_\infty(\la_0)$
as $E_\infty(\la_0) - E_\infty(\la_0 - 1)$,
this way we get the limits $E_\infty$ in Table~\ref{tab:e2at}
seemingly converged to all digits given.
With these at hand,
we can try to find a simple and practical two-point
extrapolation formula of the kind
\begin{equation}
E_\infty \approx E_\la + \left( E_\la - E_{\la - 1} \right) c_\la
\end{equation}
that would follow if it would hold that
\begin{equation}
\label{eq:eab}
E_\la \approx E_\infty + A b_\la ,
\end{equation}
with the universal $b_\la$, so that
\begin{equation}
c_\la = b_\la / (b_{\la - 1} - b_\la).
\end{equation}
Computing
\begin{equation}
\tilde{c}_\la = (E_\infty - E_\la)/( E_\la - E_{\la - 1} )
\approx c_\la
\end{equation}
over the data set of Table~\ref{tab:e2at},
we see that $\tilde{c}_\la$ are weakly system dependent,
and a conservative approximation to Eq.~(\ref{eq:eab})
is simply the shortest form of Eq.~(\ref{eq:ean}),
\begin{equation}
\label{eq:ea2}
E_\la \approx E_\infty + A/\left(\la + \frct32 \right)^3,
\end{equation}
and thus
\begin{equation}
\label{eq:ca2}
c_\la = 1 \left/ \left(\bigl(1 + 1/(\la + \frct12 ) \bigr)^3 - 1\right) \right. .
\end{equation}
We have tried to find better extrapolation formulas
by splitting the two-electron correlation energy
into its spin components ---
the same-spin part is known to have $A_3=0$ in Eq.~(\ref{eq:ean}) ---
but could not get a higher overall accuracy.
The CCSD correlation energy can as well be extrapolated
in the same way --- we have found it to be of no help
to split it into the MP2 and higher-order parts,
as the latter does not seem to show a regular behavior,
at least for smaller $\la$.
The perturbative triples energy of CCSD(T), however,
can be extrapolated well enough using the fourth power
instead of the third in Eqs.~(\ref{eq:ea2}) and~(\ref{eq:ca2}).

\begingroup
\squeezetable
\begin{table}[h]
\caption{\label{tab:e2mol}MP2 calculations on H$_2$, N$_2$ and LiF molecules.}
\begin{ruledtabular}
\begin{tabular}{lrrrrrr}
& \multicolumn{2}{c}{H$_2$} &
  \multicolumn{2}{c}{N$_2$} &
  \multicolumn{2}{c}{LiF} \\
\cline{2-3}
\cline{4-5}
\cline{6-7}
set &
 \multicolumn{1}{c}{$r$} & \multicolumn{1}{c}{$\Delta E$} &
 \multicolumn{1}{c}{$r$} & \multicolumn{1}{c}{$\Delta E$} &
 \multicolumn{1}{c}{$r$} & \multicolumn{1}{c}{$\Delta E$} \\
\hline
\texttt{L1\_1}  & 1.40946 &    -0.158356 & 2.10875 &    -0.331331 & 3.00521 &    -0.196257 \\
\texttt{L1\_2}  & 1.40515 &    -0.159956 & 2.10557 &    -0.332821 & 2.99342 &    -0.200221 \\
\texttt{L1\_3}  & 1.40570 &    -0.160033 & 2.10538 &    -0.334562 & 2.99237 &    -0.200671 \\
\texttt{L1\_4}  & 1.40542 &    -0.160064 & 2.10522 &    -0.334437 & 2.99248 &    -0.200624 \\
\texttt{L1\_5}  & 1.40547 &    -0.160064 & 2.10518 &    -0.334465 & 2.99246 &    -0.200632 \\
\texttt{L1\_6}  & 1.40544 &    -0.160065 & 2.10519 &    -0.334463 & 2.99247 &    -0.200628 \\[1pt]
\texttt{L2\_1}  & 1.39591 &    -0.164106 & 2.10430 &    -0.360115 & 2.96411 &    -0.222162 \\
\texttt{L2\_2}  & 1.39709 &    -0.164568 & 2.10216 &    -0.363413 & 2.96782 &    -0.222191 \\
\texttt{L2\_3}  & 1.39655 &    -0.164631 & 2.10234 &    -0.363672 & 2.96864 &    -0.222447 \\
\texttt{L2\_4}  & 1.39655 &    -0.164634 & 2.10231 &    -0.363682 & 2.96868 &    -0.222437 \\
\texttt{L2\_5}  & 1.39655 &    -0.164634 & 2.10230 &    -0.363683 & 2.96865 &    -0.222441 \\
                &         &\it -0.167002 &         &\it -0.372687 &         &\it -0.229689 \\[1pt]
\texttt{L3\_1}  & 1.39231 &    -0.166376 & 2.10009 &    -0.372403 & 2.96592 &    -0.228579 \\
\texttt{L3\_2}  & 1.39171 &    -0.166582 & 2.09910 &    -0.373400 & 2.96589 &    -0.228890 \\
\texttt{L3\_3}  & 1.39171 &    -0.166591 & 2.09901 &    -0.373481 & 2.96599 &    -0.228928 \\
\texttt{L3\_4}  & 1.39167 &    -0.166595 & 2.09898 &    -0.373489 & 2.96595 &    -0.228923 \\
\texttt{L3\_5}  & 1.39167 &    -0.166596 & 2.09898 &    -0.373489 & 2.96596 &    -0.228924 \\
                &         &\it -0.167683 &         &\it -0.381103 &         &\it -0.233453 \\[1pt]
\texttt{L4\_1}  & 1.39162 &    -0.167069 & 2.09743 &    -0.378245 & 2.96702 &    -0.230973 \\
\texttt{L4\_2}  & 1.39077 &    -0.167214 & 2.09678 &    -0.378617 & 2.96673 &    -0.231125 \\
\texttt{L4\_3}  & 1.39061 &    -0.167236 & 2.09667 &    -0.378723 & 2.96680 &    -0.231156 \\
\texttt{L4\_4}  & 1.39058 &    -0.167241 & 2.09665 &    -0.378735 & 2.96681 &    -0.231156 \\
\texttt{L4\_5}  & 1.39057 &    -0.167243 & 2.09664 &    -0.378735 & 2.96682 &    -0.231156 \\
                &         &\it -0.167828 &         &\it -0.383306 &         &\it -0.233674 \\[1pt]
\texttt{L5\_1}  & 1.39076 &    -0.167457 & 2.09670 &    -0.380447 & 2.96736 &    -0.232065 \\
\texttt{L5\_2}  & 1.39066 &    -0.167511 & 2.09643 &    -0.380845 & 2.96728 &    -0.232148 \\
\texttt{L5\_3}  & 1.39057 &    -0.167528 & 2.09636 &    -0.380927 & 2.96737 &    -0.232153 \\
\texttt{L5\_4}  & 1.39056 &    -0.167530 & 2.09635 &    -0.380932 & 2.96737 &    -0.232153 \\
                &         &\it -0.167878 &         &\it -0.383787 &         &\it -0.233665 \\[1pt]
\texttt{L6\_4}  & 1.39050 &    -0.167662 &         &              &         &              \\[1pt]
\texttt{L1a\_1} & 1.40587 &    -0.160673 & 2.10630 &    -0.347577 & 3.03746 &    -0.216646 \\
\texttt{L1a\_2} & 1.40370 &    -0.161083 & 2.10698 &    -0.349407 & 3.03451 &    -0.218875 \\
\texttt{L1a\_3} & 1.40402 &    -0.161151 & 2.10720 &    -0.350187 & 3.03091 &    -0.219415 \\
\texttt{L1a\_4} & 1.40378 &    -0.161191 & 2.10655 &    -0.350279 & 3.03031 &    -0.219501 \\
\texttt{L1a\_5} & 1.40379 &    -0.161189 & 2.10656 &    -0.350341 & 3.03027 &    -0.219518 \\
\texttt{L1a\_6} & 1.40378 &    -0.161192 & 2.10655 &    -0.350332 & 3.03026 &    -0.219522 \\[1pt]
\texttt{L2a\_1} & 1.39461 &    -0.165186 & 2.10393 &    -0.366668 & 2.98489 &    -0.226163 \\
\texttt{L2a\_2} & 1.39456 &    -0.165405 & 2.10172 &    -0.368966 & 2.98487 &    -0.226740 \\
\texttt{L2a\_3} & 1.39429 &    -0.165481 & 2.10110 &    -0.369448 & 2.98329 &    -0.227054 \\
\texttt{L2a\_4} & 1.39429 &    -0.165485 & 2.10101 &    -0.369488 & 2.98325 &    -0.227055 \\
\texttt{L2a\_5} & 1.39429 &    -0.165486 & 2.10100 &    -0.369496 & 2.98325 &    -0.227056 \\
                &         &\it -0.167475 &         &\it -0.377247 &         &\it -0.230105 \\[1pt]
\texttt{L3a\_1} & 1.39201 &    -0.166668 & 2.09902 &    -0.376312 & 2.97257 &    -0.230201 \\
\texttt{L3a\_2} & 1.39162 &    -0.166796 & 2.09814 &    -0.377095 & 2.97208 &    -0.230632 \\
\texttt{L3a\_3} & 1.39106 &    -0.166855 & 2.09806 &    -0.377253 & 2.97171 &    -0.230714 \\
\texttt{L3a\_4} & 1.39107 &    -0.166860 & 2.09803 &    -0.377297 & 2.97174 &    -0.230718 \\
\texttt{L3a\_5} & 1.39105 &    -0.166863 & 2.09802 &    -0.377296 & 2.97173 &    -0.230718 \\
                &         &\it -0.167770 &         &\it -0.383046 &         &\it -0.233407 \\[1pt]
\texttt{L4a\_1} & 1.39101 &    -0.167268 & 2.09721 &    -0.379951 & 2.96997 &    -0.231756 \\
\texttt{L4a\_2} & 1.39079 &    -0.167336 & 2.09714 &    -0.380194 & 2.96967 &    -0.231939 \\
\texttt{L4a\_3} & 1.39071 &    -0.167370 & 2.09706 &    -0.380318 & 2.96962 &    -0.231978 \\
\texttt{L4a\_4} & 1.39064 &    -0.167378 & 2.09702 &    -0.380352 & 2.96961 &    -0.231983 \\
\texttt{L4a\_5} & 1.39064 &    -0.167379 & 2.09702 &    -0.380354 & 2.96961 &    -0.231983 \\
                &         &\it -0.167876 &         &\it -0.383797 &         &\it -0.233438 \\[1pt]
\texttt{L5a\_1} & 1.39058 &    -0.167545 & 2.09672 &    -0.381608 & 2.96891 &    -0.232458 \\
\texttt{L5a\_2} & 1.39057 &    -0.167569 & 2.09663 &    -0.381728 & 2.96893 &    -0.232545 \\
\texttt{L5a\_3} & 1.39058 &    -0.167580 & 2.09659 &    -0.381796 & 2.96888 &    -0.232565 \\
\texttt{L5a\_4} & 1.39056 &    -0.167584 & 2.09659 &    -0.381807 & 2.96888 &    -0.232566 \\
                &         &\it -0.167889 &         &\it -0.383981 &         &\it -0.233446 
\end{tabular}
\begin{flushleft}
The bond lengths $r$ and the binding energies $\Delta E$ are in au,\\
the values in italics are extrapolated
from those on the line above.
\end{flushleft}
\end{ruledtabular}
\end{table}
\endgroup

Molecular tests in Tables~\ref{tab:e2mol} and~\ref{tab:e2dim}
show the convergence with respect
to both the number of basis functions and the quality of their approximation
with the underlying primitive set size.
The three molecules, H$_2$, N$_2$, and LiF, are prototypical
for covalent and ionic bonding and,
because of the very regular and consistent structure of our basis sets
across the periodic table, Table~\ref{tab:e2mol}
can \textit{guide the choice} of the right set for a given application.
At the other end, the weakest bonding
in the noble gas dimers He$_2$ and Ne$_2$
studied in Table~\ref{tab:e2dim}
should be under\-stood as the worst case performance ---
we see here that, without the diffuse functions,
the bond lengths and energies do converge but too slowly,
and our diffuse sets help to recover the most part
of the attractive interaction
that is somewhat overestimated and can be (over)corrected
by the counterpoise~\cite{BB70} method
so that the two binding energies seem to bracket the ``exact'' value.

\begingroup
\squeezetable
\begin{table}[h]
\caption{\label{tab:e2dim}MP2 calculations on He$_2$ and Ne$_2$ dimers.}
\begin{ruledtabular}
\begin{tabular}{lrrrrrr}
& \multicolumn{3}{c}{He$_2$} &
  \multicolumn{3}{c}{Ne$_2$} \\
\cline{2-4}
\cline{5-7}
set &
 \multicolumn{1}{c}{$r$} & \multicolumn{2}{c}{$\Delta E$} &
 \multicolumn{1}{c}{$r$} & \multicolumn{2}{c}{$\Delta E$} \\
\hline
 \texttt{L1\_6}  & 7.1087 &     -1.43 &     -1.43 & 7.3415 &     -5.59 &      -5.19 \\
 \texttt{L2\_5}  & 6.5483 &     -4.81 &     -4.81 & 6.6842 &    -21.93 &     -19.50 \\
 \texttt{L3\_5}  & 6.2747 &     -8.35 &     -8.35 & 6.4521 &    -36.40 &     -33.76 \\
 \texttt{L4\_5}  & 6.1112 &    -11.41 &    -11.41 & 6.3184 &    -48.17 &     -45.86 \\
 \texttt{L5\_4}  & 6.0039 &    -13.92 &    -13.91 & 6.2362 &    -56.85 &     -55.05 \\
 \texttt{L6\_4}  & 5.9309 &    -15.93 &    -15.91 & 6.1793 &    -63.69 &     -62.32 \\
 \texttt{L7\_4}  & 5.8808 &    -17.52 &    -17.50 &        &           &            \\[1pt]
 \texttt{L1a\_1} & 6.0461 &    -16.78 &    -11.85 & 6.1093 &   -114.72 &     -42.32 \\
 \texttt{L1a\_2} & 5.9113 &    -17.28 &    -10.79 & 5.7486 &   -184.44 &      -6.93 \\
 \texttt{L1a\_3} & 5.8329 &    -21.95 &     -9.72 & 5.6760 &   -270.77 &      +7.48 \\
 \texttt{L1a\_4} & 5.8239 &    -23.15 &     -9.60 & 5.6675 &   -288.64 &      +9.13 \\
 \texttt{L1a\_5} & 5.8228 &    -23.49 &     -9.58 & 5.6682 &   -291.92 &      +9.04 \\
 \texttt{L1a\_6} & 5.8224 &    -23.58 &     -9.57 & 5.6690 &   -292.01 &      +8.82 \\[1pt]
 \texttt{L2a\_1} & 5.8699 &    -18.63 &    -17.00 & 5.9650 &    -96.94 &     -62.23 \\
 \texttt{L2a\_2} & 5.8507 &    -19.82 &    -16.97 & 5.9033 &   -139.10 &     -58.32 \\
 \texttt{L2a\_3} & 5.8384 &    -20.62 &    -16.70 & 5.8724 &   -157.45 &     -56.09 \\
 \texttt{L2a\_4} & 5.8374 &    -20.77 &    -16.70 & 5.8700 &   -159.35 &     -55.94 \\
 \texttt{L2a\_5} & 5.8377 &    -20.80 &    -16.69 & 5.8698 &   -159.72 &     -55.91 \\
                 &        &\it -19.20 &\it -20.63 &        &\it -90.09 &\it  -71.61 \\[1pt]
 \texttt{L3a\_1} & 5.8133 &    -20.50 &    -19.12 & 6.0311 &    -90.36 &     -73.03 \\
 \texttt{L3a\_2} & 5.8298 &    -20.18 &    -18.97 & 6.0138 &   -101.21 &     -72.23 \\
 \texttt{L3a\_3} & 5.8253 &    -20.32 &    -18.88 & 5.9976 &   -104.79 &     -71.97 \\
 \texttt{L3a\_4} & 5.8236 &    -20.42 &    -18.89 & 5.9932 &   -106.00 &     -71.95 \\
 \texttt{L3a\_5} & 5.8240 &    -20.43 &    -18.88 & 5.9933 &   -106.22 &     -71.94 \\
                 &        &\it -20.10 &\it -20.88 &        &\it -60.76 &\it  -79.39 \\[1pt]
 \texttt{L4a\_1} & 5.8031 &    -21.00 &    -20.40 & 6.0535 &    -87.59 &     -78.30 \\
 \texttt{L4a\_2} & 5.8117 &    -20.71 &    -20.11 & 6.0430 &    -89.68 &     -77.60 \\
 \texttt{L4a\_3} & 5.8090 &    -20.79 &    -20.10 & 6.0339 &    -91.17 &     -77.35 \\
 \texttt{L4a\_4} & 5.8077 &    -20.84 &    -20.12 & 6.0314 &    -91.78 &     -77.38 \\
 \texttt{L4a\_5} & 5.8078 &    -20.84 &    -20.12 & 6.0315 &    -91.89 &     -77.37 \\
                 &        &\it -21.35 &\it -21.66 &        &\it -74.72 &\it  -82.83 \\[1pt]
 \texttt{L5a\_1} & 5.7920 &    -21.82 &    -21.15 & 6.0451 &    -87.45 &     -81.07 \\
 \texttt{L5a\_2} & 5.7985 &    -21.31 &    -20.91 & 6.0440 &    -87.46 &     -80.32 \\
 \texttt{L5a\_3} & 5.7966 &    -21.32 &    -20.93 & 6.0407 &    -87.71 &     -80.47 \\
 \texttt{L5a\_4} & 5.7963 &    -21.33 &    -20.94 & 6.0401 &    -87.84 &     -80.49 \\
                 &        &\it -22.09 &\it -22.22 &        &\it -81.62 &\it  -85.11 \\[1pt]
 \texttt{L6a\_4} & 5.7893 &    -21.73 &    -21.49 &        &           &            
\end{tabular}
\begin{flushleft}
The binding energies $\Delta E$ are in microhartrees and computed\\
either relative to the isolated atoms (first column)\\
or with the counterpoise~\cite{BB70} correction (second column), \\
the values in italics are extrapolated
from those on the line above. \\
The bond lengths $r$ are in bohrs.
\end{flushleft}
\end{ruledtabular}
\end{table}
\endgroup

The extrapolation of the correlation energy
at least does not hurt these weakest bonds,
leading to some underbinding,
but is a great improvement for the strong chemical bonds
as seen in Table~\ref{tab:e2mol},
so it should be helpful for all molecular systems.

The dipole polarizabilities in Table~\ref{tab:pol}
clearly witness the need for the diffuse functions,
using Eq.~(\ref{eq:Ed}) we add one more ``\texttt{b}''-set
in \texttt{L}$\la$\texttt{ab} which yields
the highest accuracy, but the \texttt{L}$\la$\texttt{a}
are already quite good and should be used
to get the accurate intermolecular interactions.

\begingroup
\squeezetable
\begin{table}[h]
\caption{\label{tab:pol}MP2 polarizabilities (au) of atoms and molecules.}
\begin{ruledtabular}
\begin{tabular}{lrrrrrr}
set &
 \multicolumn{1}{c}{He} &
 \multicolumn{1}{c}{Ne} &
 \multicolumn{2}{c}{H$_2$} &
 \multicolumn{2}{c}{N$_2$} \\
\hline
\texttt{L1\_6}   & 0.367 & 0.618 & 6.280 & 2.220 & 11.529 &  5.595 \\
\texttt{L2\_5}   & 0.675 & 1.154 & 6.430 & 3.320 & 12.666 &  7.277 \\
\texttt{L3\_5}   & 0.886 & 1.529 & 6.405 & 3.840 & 13.478 &  8.295 \\
\texttt{L4\_5}   & 1.023 & 1.802 & 6.430 & 4.114 & 13.964 &  8.892 \\
\texttt{L5\_4}   & 1.113 & 1.990 & 6.398 & 4.284 & 14.147 &  9.265 \\[1pt]
\texttt{L1a\_6}  & 1.203 & 2.132 & 6.577 & 4.477 & 14.498 &  9.676 \\
\texttt{L2a\_5}  & 1.252 & 2.331 & 6.364 & 4.486 & 14.380 &  9.897 \\
\texttt{L3a\_5}  & 1.283 & 2.445 & 6.377 & 4.500 & 14.461 & 10.018 \\
\texttt{L4a\_5}  & 1.304 & 2.517 & 6.360 & 4.523 & 14.458 & 10.072 \\
\texttt{L5a\_4}  & 1.318 & 2.560 & 6.351 & 4.535 & 14.461 & 10.105 \\[1pt]
\texttt{L1ab\_6} & 1.354 & 2.711 & 6.569 & 4.630 & 14.658 & 10.220 \\
\texttt{L2ab\_5} & 1.360 & 2.712 & 6.395 & 4.547 & 14.588 & 10.239 \\
\texttt{L3ab\_5} & 1.361 & 2.704 & 6.359 & 4.551 & 14.517 & 10.200 \\
\texttt{L4ab\_5} & 1.360 & 2.701 & 6.351 & 4.557 & 14.487 & 10.191
\end{tabular}
\end{ruledtabular}
\end{table}
\endgroup

The minimal primitive representation
of angular polarization functions of our
\texttt{L}$\la$\texttt{\_1} and \texttt{L}$\la$\texttt{a\_1}
sets for lighter atoms
can be compared one-to-one
with that of the classical works~\cite{D89,KDH92,WD93},
a remarkably close match
of the exponents for atoms He, B--Ne, and Al-Ar
can be seen (less so for H as we take the atom and not the
H$_2$ molecule),
and this must be a very good sign for us all.

\section{Conclusions}

The atomic basis sets are now made available~\cite{SM}
that allow quantitative calculations of the structure and energetics
of typical molecular systems.
We are somewhat sorry for our belated report,
for they have already been, and are being, used by our colleagues
in their studies of \textit{unusual molecules}
under the unusual conditions of low-temperature high-energy
chemistry~\cite{SSTF17,RTF17,RTFK17,KTF17,KTNF16,STF16,SSTF18}.

They can also be used to set up
a database of accurate reference values
for the parametrization of electronic structure models~\cite{L11,B17}
or molecular mechanical force fields of all kinds.
We are looking forward to their further useful applications
for the good of chemistry, chemists, and society.
Atoms are many, molecules are endless,
but we are few, not to say alone.


\begin{thebibliography}{102}%
\makeatletter
\providecommand \@ifxundefined [1]{%
 \@ifx{#1\undefined}
}%
\providecommand \@ifnum [1]{%
 \ifnum #1\expandafter \@firstoftwo
 \else \expandafter \@secondoftwo
 \fi
}%
\providecommand \@ifx [1]{%
 \ifx #1\expandafter \@firstoftwo
 \else \expandafter \@secondoftwo
 \fi
}%
\providecommand \natexlab [1]{#1}%
\providecommand \enquote  [1]{``#1''}%
\providecommand \bibnamefont  [1]{#1}%
\providecommand \bibfnamefont [1]{#1}%
\providecommand \citenamefont [1]{#1}%
\providecommand \href@noop [0]{\@secondoftwo}%
\providecommand \href [0]{\begingroup \@sanitize@url \@href}%
\providecommand \@href[1]{\@@startlink{#1}\@@href}%
\providecommand \@@href[1]{\endgroup#1\@@endlink}%
\providecommand \@sanitize@url [0]{\catcode `\\12\catcode `\$12\catcode
  `\&12\catcode `\#12\catcode `\^12\catcode `\_12\catcode `\%12\relax}%
\providecommand \@@startlink[1]{}%
\providecommand \@@endlink[0]{}%
\providecommand \url  [0]{\begingroup\@sanitize@url \@url }%
\providecommand \@url [1]{\endgroup\@href {#1}{\urlprefix }}%
\providecommand \urlprefix  [0]{URL }%
\providecommand \Eprint [0]{\href }%
\providecommand \doibase [0]{http://dx.doi.org/}%
\providecommand \selectlanguage [0]{\@gobble}%
\providecommand \bibinfo  [0]{\@secondoftwo}%
\providecommand \bibfield  [0]{\@secondoftwo}%
\providecommand \translation [1]{[#1]}%
\providecommand \BibitemOpen [0]{}%
\providecommand \bibitemStop [0]{}%
\providecommand \bibitemNoStop [0]{.\EOS\space}%
\providecommand \EOS [0]{\spacefactor3000\relax}%
\providecommand \BibitemShut  [1]{\csname bibitem#1\endcsname}%
\let\auto@bib@innerbib\@empty
\bibitem [{\citenamefont {Heitler}\ and\ \citenamefont {London}(1927)}]{HL27}%
  \BibitemOpen
  \bibfield  {author} {\bibinfo {author} {\bibfnamefont {W.}~\bibnamefont
  {Heitler}}\ and\ \bibinfo {author} {\bibfnamefont {F.}~\bibnamefont
  {London}},\ }\href {\doibase 10.1007/bf01397394} {\bibfield  {journal}
  {\bibinfo  {journal} {Z. Physik}\ }\textbf {\bibinfo {volume} {44}},\
  \bibinfo {pages} {455} (\bibinfo {year} {1927})}\BibitemShut {NoStop}%
\bibitem [{\citenamefont {Lennard-Jones}(1929)}]{L29}%
  \BibitemOpen
  \bibfield  {author} {\bibinfo {author} {\bibfnamefont {J.~E.}\ \bibnamefont
  {Lennard-Jones}},\ }\href {\doibase 10.1039/tf9292500668} {\bibfield
  {journal} {\bibinfo  {journal} {Trans. Faraday Soc.}\ }\textbf {\bibinfo
  {volume} {25}},\ \bibinfo {pages} {668} (\bibinfo {year} {1929})}\BibitemShut
  {NoStop}%
\bibitem [{\citenamefont {Schr\"odinger}(1926{\natexlab{a}})}]{S26}%
  \BibitemOpen
  \bibfield  {author} {\bibinfo {author} {\bibfnamefont {E.}~\bibnamefont
  {Schr\"odinger}},\ }\href@noop {} {\bibfield  {journal} {\bibinfo  {journal}
  {Ann. Phys.}\ }\textbf {\bibinfo {volume} {79}},\ \bibinfo {pages} {361}
  (\bibinfo {year} {1926}{\natexlab{a}})}\BibitemShut {NoStop}%
\bibitem [{\citenamefont {Schr\"odinger}(1926{\natexlab{b}})}]{S26b}%
  \BibitemOpen
  \bibfield  {author} {\bibinfo {author} {\bibfnamefont {E.}~\bibnamefont
  {Schr\"odinger}},\ }\href {\doibase 10.1103/PhysRev.28.1049} {\bibfield
  {journal} {\bibinfo  {journal} {Phys. Rev.}\ }\textbf {\bibinfo {volume}
  {28}},\ \bibinfo {pages} {1049} (\bibinfo {year}
  {1926}{\natexlab{b}})}\BibitemShut {NoStop}%
\bibitem [{\citenamefont {Dirac}(1928)}]{D28}%
  \BibitemOpen
  \bibfield  {author} {\bibinfo {author} {\bibfnamefont {P.~A.~M.}\
  \bibnamefont {Dirac}},\ }\href {\doibase 10.1098/rspa.1928.0023} {\bibfield
  {journal} {\bibinfo  {journal} {Proc. R. Soc. Lon. Ser. A}\ }\textbf
  {\bibinfo {volume} {117}},\ \bibinfo {pages} {610} (\bibinfo {year}
  {1928})}\BibitemShut {NoStop}%
\bibitem [{\citenamefont {Zener}(1930)}]{Z30}%
  \BibitemOpen
  \bibfield  {author} {\bibinfo {author} {\bibfnamefont {C.}~\bibnamefont
  {Zener}},\ }\href {\doibase 10.1103/PhysRev.36.51} {\bibfield  {journal}
  {\bibinfo  {journal} {Phys. Rev.}\ }\textbf {\bibinfo {volume} {36}},\
  \bibinfo {pages} {51} (\bibinfo {year} {1930})}\BibitemShut {NoStop}%
\bibitem [{\citenamefont {Slater}(1930)}]{S30}%
  \BibitemOpen
  \bibfield  {author} {\bibinfo {author} {\bibfnamefont {J.~C.}\ \bibnamefont
  {Slater}},\ }\href {\doibase 10.1103/PhysRev.36.57} {\bibfield  {journal}
  {\bibinfo  {journal} {Phys. Rev.}\ }\textbf {\bibinfo {volume} {36}},\
  \bibinfo {pages} {57} (\bibinfo {year} {1930})}\BibitemShut {NoStop}%
\bibitem [{\citenamefont {Hartree}(1928)}]{H28}%
  \BibitemOpen
  \bibfield  {author} {\bibinfo {author} {\bibfnamefont {D.~R.}\ \bibnamefont
  {Hartree}},\ }\href@noop {} {\bibfield  {journal} {\bibinfo  {journal} {P.
  Camb. Philos. Soc.}\ }\textbf {\bibinfo {volume} {24}},\ \bibinfo {pages}
  {89} (\bibinfo {year} {1928})}\BibitemShut {NoStop}%
\bibitem [{\citenamefont {Fock}(1930)}]{F30}%
  \BibitemOpen
  \bibfield  {author} {\bibinfo {author} {\bibfnamefont {V.}~\bibnamefont
  {Fock}},\ }\href {\doibase 10.1007/BF01340294} {\bibfield  {journal}
  {\bibinfo  {journal} {Z. Physik}\ }\textbf {\bibinfo {volume} {61}},\
  \bibinfo {pages} {126} (\bibinfo {year} {1930})}\BibitemShut {NoStop}%
\bibitem [{\citenamefont {Roothaan}(1951)}]{R51}%
  \BibitemOpen
  \bibfield  {author} {\bibinfo {author} {\bibfnamefont {C.~C.~J.}\
  \bibnamefont {Roothaan}},\ }\href {\doibase 10.1103/RevModPhys.23.69}
  {\bibfield  {journal} {\bibinfo  {journal} {Rev. Mod. Phys.}\ }\textbf
  {\bibinfo {volume} {23}},\ \bibinfo {pages} {69} (\bibinfo {year}
  {1951})}\BibitemShut {NoStop}%
\bibitem [{\citenamefont {L\"owdin}(1955)}]{L55}%
  \BibitemOpen
  \bibfield  {author} {\bibinfo {author} {\bibfnamefont {P.-O.}\ \bibnamefont
  {L\"owdin}},\ }\href {\doibase 10.1103/PhysRev.97.1474} {\bibfield  {journal}
  {\bibinfo  {journal} {Phys. Rev.}\ }\textbf {\bibinfo {volume} {97}},\
  \bibinfo {pages} {1474} (\bibinfo {year} {1955})}\BibitemShut {NoStop}%
\bibitem [{\citenamefont {Kohn}\ and\ \citenamefont {Sham}(1965)}]{KS65}%
  \BibitemOpen
  \bibfield  {author} {\bibinfo {author} {\bibfnamefont {W.}~\bibnamefont
  {Kohn}}\ and\ \bibinfo {author} {\bibfnamefont {L.~J.}\ \bibnamefont
  {Sham}},\ }\href {\doibase 10.1103/PhysRev.140.A1133} {\bibfield  {journal}
  {\bibinfo  {journal} {Phys. Rev.}\ }\textbf {\bibinfo {volume} {140}},\
  \bibinfo {pages} {A1133} (\bibinfo {year} {1965})}\BibitemShut {NoStop}%
\bibitem [{\citenamefont {Baerends}, \citenamefont {Ellis},\ and\ \citenamefont
  {Ros}(1973)}]{BER73}%
  \BibitemOpen
  \bibfield  {author} {\bibinfo {author} {\bibfnamefont {E.~J.}\ \bibnamefont
  {Baerends}}, \bibinfo {author} {\bibfnamefont {D.~E.}\ \bibnamefont {Ellis}},
  \ and\ \bibinfo {author} {\bibfnamefont {P.}~\bibnamefont {Ros}},\ }\href
  {\doibase 10.1016/0301-0104(73)80059-X} {\bibfield  {journal} {\bibinfo
  {journal} {Chem. Phys.}\ }\textbf {\bibinfo {volume} {2}},\ \bibinfo {pages}
  {41} (\bibinfo {year} {1973})}\BibitemShut {NoStop}%
\bibitem [{\citenamefont {Lenthe}\ and\ \citenamefont {Baerends}(2003)}]{LB03}%
  \BibitemOpen
  \bibfield  {author} {\bibinfo {author} {\bibfnamefont {E.~V.}\ \bibnamefont
  {Lenthe}}\ and\ \bibinfo {author} {\bibfnamefont {E.~J.}\ \bibnamefont
  {Baerends}},\ }\href {\doibase 10.1002/jcc.10255} {\bibfield  {journal}
  {\bibinfo  {journal} {J. Comp. Chem.}\ }\textbf {\bibinfo {volume} {24}},\
  \bibinfo {pages} {1142} (\bibinfo {year} {2003})}\BibitemShut {NoStop}%
\bibitem [{\citenamefont {van Lenthe}, \citenamefont {Baerends},\ and\
  \citenamefont {Snijders}(1993)}]{LBS93}%
  \BibitemOpen
  \bibfield  {author} {\bibinfo {author} {\bibfnamefont {E.}~\bibnamefont {van
  Lenthe}}, \bibinfo {author} {\bibfnamefont {E.~J.}\ \bibnamefont {Baerends}},
  \ and\ \bibinfo {author} {\bibfnamefont {J.~G.}\ \bibnamefont {Snijders}},\
  }\href {\doibase 10.1063/1.466059} {\bibfield  {journal} {\bibinfo  {journal}
  {J. Chem. Phys.}\ }\textbf {\bibinfo {volume} {99}},\ \bibinfo {pages} {4597}
  (\bibinfo {year} {1993})}\BibitemShut {NoStop}%
\bibitem [{\citenamefont {Boys}(1950)}]{B50}%
  \BibitemOpen
  \bibfield  {author} {\bibinfo {author} {\bibfnamefont {S.~F.}\ \bibnamefont
  {Boys}},\ }\href {\doibase 10.1098/rspa.1950.0036} {\bibfield  {journal}
  {\bibinfo  {journal} {Proc. R. Soc. A}\ }\textbf {\bibinfo {volume} {200}},\
  \bibinfo {pages} {542} (\bibinfo {year} {1950})}\BibitemShut {NoStop}%
\bibitem [{\citenamefont {Hehre}, \citenamefont {Stewart},\ and\ \citenamefont
  {Pople}(1969)}]{HSP69}%
  \BibitemOpen
  \bibfield  {author} {\bibinfo {author} {\bibfnamefont {W.~J.}\ \bibnamefont
  {Hehre}}, \bibinfo {author} {\bibfnamefont {R.~F.}\ \bibnamefont {Stewart}},
  \ and\ \bibinfo {author} {\bibfnamefont {J.~A.}\ \bibnamefont {Pople}},\
  }\href {\doibase 10.1063/1.1672392} {\bibfield  {journal} {\bibinfo
  {journal} {J. Chem. Phys.}\ }\textbf {\bibinfo {volume} {51}},\ \bibinfo
  {pages} {2657} (\bibinfo {year} {1969})}\BibitemShut {NoStop}%
\bibitem [{\citenamefont {Stewart}(1970)}]{S70}%
  \BibitemOpen
  \bibfield  {author} {\bibinfo {author} {\bibfnamefont {R.~F.}\ \bibnamefont
  {Stewart}},\ }\href {\doibase 10.1063/1.1672702} {\bibfield  {journal}
  {\bibinfo  {journal} {J. Chem. Phys.}\ }\textbf {\bibinfo {volume} {52}},\
  \bibinfo {pages} {431} (\bibinfo {year} {1970})}\BibitemShut {NoStop}%
\bibitem [{\citenamefont {Ditchfield}, \citenamefont {Hehre},\ and\
  \citenamefont {Pople}(1970)}]{DHP70}%
  \BibitemOpen
  \bibfield  {author} {\bibinfo {author} {\bibfnamefont {R.}~\bibnamefont
  {Ditchfield}}, \bibinfo {author} {\bibfnamefont {W.~J.}\ \bibnamefont
  {Hehre}}, \ and\ \bibinfo {author} {\bibfnamefont {J.~A.}\ \bibnamefont
  {Pople}},\ }\href {\doibase 10.1063/1.1672736} {\bibfield  {journal}
  {\bibinfo  {journal} {J. Chem. Phys.}\ }\textbf {\bibinfo {volume} {52}},\
  \bibinfo {pages} {5001} (\bibinfo {year} {1970})}\BibitemShut {NoStop}%
\bibitem [{\citenamefont {Ditchfield}, \citenamefont {Hehre},\ and\
  \citenamefont {Pople}(1971)}]{DHP71}%
  \BibitemOpen
  \bibfield  {author} {\bibinfo {author} {\bibfnamefont {R.}~\bibnamefont
  {Ditchfield}}, \bibinfo {author} {\bibfnamefont {W.~J.}\ \bibnamefont
  {Hehre}}, \ and\ \bibinfo {author} {\bibfnamefont {J.~A.}\ \bibnamefont
  {Pople}},\ }\href {\doibase 10.1063/1.1674902} {\bibfield  {journal}
  {\bibinfo  {journal} {J. Chem. Phys.}\ }\textbf {\bibinfo {volume} {54}},\
  \bibinfo {pages} {724} (\bibinfo {year} {1971})}\BibitemShut {NoStop}%
\bibitem [{\citenamefont {Hehre}, \citenamefont {Ditchfield},\ and\
  \citenamefont {Pople}(1972)}]{HDP72}%
  \BibitemOpen
  \bibfield  {author} {\bibinfo {author} {\bibfnamefont {W.~J.}\ \bibnamefont
  {Hehre}}, \bibinfo {author} {\bibfnamefont {R.}~\bibnamefont {Ditchfield}}, \
  and\ \bibinfo {author} {\bibfnamefont {J.~A.}\ \bibnamefont {Pople}},\ }\href
  {\doibase 10.1063/1.1677527} {\bibfield  {journal} {\bibinfo  {journal} {J.
  Chem. Phys.}\ }\textbf {\bibinfo {volume} {56}},\ \bibinfo {pages} {2257}
  (\bibinfo {year} {1972})}\BibitemShut {NoStop}%
\bibitem [{\citenamefont {Hariharan}\ and\ \citenamefont {Pople}(1973)}]{HP73}%
  \BibitemOpen
  \bibfield  {author} {\bibinfo {author} {\bibfnamefont {P.~C.}\ \bibnamefont
  {Hariharan}}\ and\ \bibinfo {author} {\bibfnamefont {J.~A.}\ \bibnamefont
  {Pople}},\ }\href {\doibase 10.1007/BF00533485} {\bibfield  {journal}
  {\bibinfo  {journal} {Theor. Chim. Acta}\ }\textbf {\bibinfo {volume} {28}},\
  \bibinfo {pages} {213} (\bibinfo {year} {1973})}\BibitemShut {NoStop}%
\bibitem [{\citenamefont {Frisch}\ and\ \citenamefont {Pople}(1984)}]{FP84}%
  \BibitemOpen
  \bibfield  {author} {\bibinfo {author} {\bibfnamefont {M.~J.}\ \bibnamefont
  {Frisch}}\ and\ \bibinfo {author} {\bibfnamefont {J.~A.}\ \bibnamefont
  {Pople}},\ }\href {\doibase 10.1063/1.447079} {\bibfield  {journal} {\bibinfo
   {journal} {J. Chem. Phys.}\ }\textbf {\bibinfo {volume} {80}},\ \bibinfo
  {pages} {3265} (\bibinfo {year} {1984})}\BibitemShut {NoStop}%
\bibitem [{\citenamefont {M{\o}ller}\ and\ \citenamefont
  {Plesset}(1934)}]{MP34}%
  \BibitemOpen
  \bibfield  {author} {\bibinfo {author} {\bibfnamefont {C.}~\bibnamefont
  {M{\o}ller}}\ and\ \bibinfo {author} {\bibfnamefont {M.~S.}\ \bibnamefont
  {Plesset}},\ }\href {\doibase 10.1103/PhysRev.46.618} {\bibfield  {journal}
  {\bibinfo  {journal} {Phys. Rev.}\ }\textbf {\bibinfo {volume} {46}},\
  \bibinfo {pages} {618} (\bibinfo {year} {1934})}\BibitemShut {NoStop}%
\bibitem [{\citenamefont {Bartlett}(1975)}]{B75}%
  \BibitemOpen
  \bibfield  {author} {\bibinfo {author} {\bibfnamefont {R.~J.}\ \bibnamefont
  {Bartlett}},\ }\href {\doibase 10.1063/1.430878} {\bibfield  {journal}
  {\bibinfo  {journal} {J. Chem. Phys.}\ }\textbf {\bibinfo {volume} {62}},\
  \bibinfo {pages} {3258} (\bibinfo {year} {1975})}\BibitemShut {NoStop}%
\bibitem [{\citenamefont {Pople}, \citenamefont {Seeger},\ and\ \citenamefont
  {Krishnan}(1977)}]{PSK77}%
  \BibitemOpen
  \bibfield  {author} {\bibinfo {author} {\bibfnamefont {J.~A.}\ \bibnamefont
  {Pople}}, \bibinfo {author} {\bibfnamefont {R.}~\bibnamefont {Seeger}}, \
  and\ \bibinfo {author} {\bibfnamefont {R.}~\bibnamefont {Krishnan}},\ }\href
  {\doibase 10.1002/qua.560120820} {\bibfield  {journal} {\bibinfo  {journal}
  {Int. J. Quantum Chem.}\ }\textbf {\bibinfo {volume} {12}},\ \bibinfo {pages}
  {149} (\bibinfo {year} {1977})}\BibitemShut {NoStop}%
\bibitem [{\citenamefont {Krishnan}\ and\ \citenamefont {Pople}(1978)}]{KP78}%
  \BibitemOpen
  \bibfield  {author} {\bibinfo {author} {\bibfnamefont {R.}~\bibnamefont
  {Krishnan}}\ and\ \bibinfo {author} {\bibfnamefont {J.~A.}\ \bibnamefont
  {Pople}},\ }\href {\doibase 10.1002/qua.560140109} {\bibfield  {journal}
  {\bibinfo  {journal} {Int. J. Quantum Chem.}\ }\textbf {\bibinfo {volume}
  {14}},\ \bibinfo {pages} {91} (\bibinfo {year} {1978})}\BibitemShut {NoStop}%
\bibitem [{\citenamefont {Krishnan}, \citenamefont {Frisch},\ and\
  \citenamefont {Pople}(1980)}]{KFP80}%
  \BibitemOpen
  \bibfield  {author} {\bibinfo {author} {\bibfnamefont {R.}~\bibnamefont
  {Krishnan}}, \bibinfo {author} {\bibfnamefont {M.~J.}\ \bibnamefont
  {Frisch}}, \ and\ \bibinfo {author} {\bibfnamefont {J.~A.}\ \bibnamefont
  {Pople}},\ }\href {\doibase 10.1063/1.439657} {\bibfield  {journal} {\bibinfo
   {journal} {J. Chem. Phys.}\ }\textbf {\bibinfo {volume} {72}},\ \bibinfo
  {pages} {4244} (\bibinfo {year} {1980})}\BibitemShut {NoStop}%
\bibitem [{\citenamefont {Raghavachari}\ \emph {et~al.}(1990)\citenamefont
  {Raghavachari}, \citenamefont {Pople}, \citenamefont {Replogle},\ and\
  \citenamefont {Head-Gordon}}]{RPRH90}%
  \BibitemOpen
  \bibfield  {author} {\bibinfo {author} {\bibfnamefont {K.}~\bibnamefont
  {Raghavachari}}, \bibinfo {author} {\bibfnamefont {J.~A.}\ \bibnamefont
  {Pople}}, \bibinfo {author} {\bibfnamefont {E.~S.}\ \bibnamefont {Replogle}},
  \ and\ \bibinfo {author} {\bibfnamefont {M.}~\bibnamefont {Head-Gordon}},\
  }\href {\doibase 10.1021/j100377a033} {\bibfield  {journal} {\bibinfo
  {journal} {J. Phys. Chem.}\ }\textbf {\bibinfo {volume} {94}},\ \bibinfo
  {pages} {5579} (\bibinfo {year} {1990})}\BibitemShut {NoStop}%
\bibitem [{\citenamefont {Coester}(1958)}]{C58}%
  \BibitemOpen
  \bibfield  {author} {\bibinfo {author} {\bibfnamefont {F.}~\bibnamefont
  {Coester}},\ }\href {\doibase 10.1016/0029-5582(58)90280-3} {\bibfield
  {journal} {\bibinfo  {journal} {Nucl. Phys.}\ }\textbf {\bibinfo {volume}
  {7}},\ \bibinfo {pages} {421} (\bibinfo {year} {1958})}\BibitemShut {NoStop}%
\bibitem [{\citenamefont {Coester}\ and\ \citenamefont
  {K\"ummel}(1960)}]{CK60}%
  \BibitemOpen
  \bibfield  {author} {\bibinfo {author} {\bibfnamefont {F.}~\bibnamefont
  {Coester}}\ and\ \bibinfo {author} {\bibfnamefont {H.}~\bibnamefont
  {K\"ummel}},\ }\href {\doibase 10.1016/0029-5582(60)90140-1} {\bibfield
  {journal} {\bibinfo  {journal} {Nucl. Phys.}\ }\textbf {\bibinfo {volume}
  {17}},\ \bibinfo {pages} {477} (\bibinfo {year} {1960})}\BibitemShut
  {NoStop}%
\bibitem [{\citenamefont {\v{C}i\v{z}ek}(1966)}]{C66}%
  \BibitemOpen
  \bibfield  {author} {\bibinfo {author} {\bibfnamefont {J.}~\bibnamefont
  {\v{C}i\v{z}ek}},\ }\href {\doibase 10.1063/1.1727484} {\bibfield  {journal}
  {\bibinfo  {journal} {J. Chem. Phys.}\ }\textbf {\bibinfo {volume} {45}},\
  \bibinfo {pages} {4256} (\bibinfo {year} {1966})}\BibitemShut {NoStop}%
\bibitem [{\citenamefont {Purvis}\ and\ \citenamefont {Bartlett}(1982)}]{PB82}%
  \BibitemOpen
  \bibfield  {author} {\bibinfo {author} {\bibfnamefont {G.~D.}\ \bibnamefont
  {Purvis}}\ and\ \bibinfo {author} {\bibfnamefont {R.~J.}\ \bibnamefont
  {Bartlett}},\ }\href {\doibase 10.1063/1.443164} {\bibfield  {journal}
  {\bibinfo  {journal} {J. Chem. Phys.}\ }\textbf {\bibinfo {volume} {76}},\
  \bibinfo {pages} {1910} (\bibinfo {year} {1982})}\BibitemShut {NoStop}%
\bibitem [{\citenamefont {Handy}\ \emph {et~al.}(1989)\citenamefont {Handy},
  \citenamefont {Pople}, \citenamefont {Head-Gordon}, \citenamefont
  {Raghavachari},\ and\ \citenamefont {Trucks}}]{HPHRT89}%
  \BibitemOpen
  \bibfield  {author} {\bibinfo {author} {\bibfnamefont {N.~C.}\ \bibnamefont
  {Handy}}, \bibinfo {author} {\bibfnamefont {J.~A.}\ \bibnamefont {Pople}},
  \bibinfo {author} {\bibfnamefont {M.}~\bibnamefont {Head-Gordon}}, \bibinfo
  {author} {\bibfnamefont {K.}~\bibnamefont {Raghavachari}}, \ and\ \bibinfo
  {author} {\bibfnamefont {G.~W.}\ \bibnamefont {Trucks}},\ }\href {\doibase
  10.1016/0009-2614(89)85013-4} {\bibfield  {journal} {\bibinfo  {journal}
  {Chem. Phys. Lett.}\ }\textbf {\bibinfo {volume} {164}},\ \bibinfo {pages}
  {185} (\bibinfo {year} {1989})}\BibitemShut {NoStop}%
\bibitem [{\citenamefont {Ragavachari}\ \emph {et~al.}(1989)\citenamefont
  {Ragavachari}, \citenamefont {Trucks}, \citenamefont {Pople},\ and\
  \citenamefont {Head-Gordon}}]{RTPH89}%
  \BibitemOpen
  \bibfield  {author} {\bibinfo {author} {\bibfnamefont {K.}~\bibnamefont
  {Ragavachari}}, \bibinfo {author} {\bibfnamefont {G.~W.}\ \bibnamefont
  {Trucks}}, \bibinfo {author} {\bibfnamefont {J.~A.}\ \bibnamefont {Pople}}, \
  and\ \bibinfo {author} {\bibfnamefont {M.}~\bibnamefont {Head-Gordon}},\
  }\href {\doibase 10.1016/S0009-2614(89)87395-6} {\bibfield  {journal}
  {\bibinfo  {journal} {Chem. Phys. Lett.}\ }\textbf {\bibinfo {volume}
  {157}},\ \bibinfo {pages} {479} (\bibinfo {year} {1989})}\BibitemShut
  {NoStop}%
\bibitem [{\citenamefont {Friesner}(1988)}]{F88}%
  \BibitemOpen
  \bibfield  {author} {\bibinfo {author} {\bibfnamefont {R.~A.}\ \bibnamefont
  {Friesner}},\ }\href {\doibase 10.1021/j100322a017} {\bibfield  {journal}
  {\bibinfo  {journal} {J. Phys. Chem.}\ }\textbf {\bibinfo {volume} {92}},\
  \bibinfo {pages} {3091} (\bibinfo {year} {1988})}\BibitemShut {NoStop}%
\bibitem [{\citenamefont {Ringnalda}, \citenamefont {Belhadj},\ and\
  \citenamefont {Friesner}(1990)}]{RBF90}%
  \BibitemOpen
  \bibfield  {author} {\bibinfo {author} {\bibfnamefont {M.~N.}\ \bibnamefont
  {Ringnalda}}, \bibinfo {author} {\bibfnamefont {M.}~\bibnamefont {Belhadj}},
  \ and\ \bibinfo {author} {\bibfnamefont {R.~A.}\ \bibnamefont {Friesner}},\
  }\href {\doibase 10.1063/1.458819} {\bibfield  {journal} {\bibinfo  {journal}
  {J. Chem. Phys.}\ }\textbf {\bibinfo {volume} {93}},\ \bibinfo {pages} {3397}
  (\bibinfo {year} {1990})}\BibitemShut {NoStop}%
\bibitem [{\citenamefont {Greeley}\ \emph {et~al.}(1994)\citenamefont
  {Greeley}, \citenamefont {Russo}, \citenamefont {Mainz}, \citenamefont
  {Friesner}, \citenamefont {Langlois}, \citenamefont {Goddard}, \citenamefont
  {Donnelly},\ and\ \citenamefont {Ringnalda}}]{GRMFLGDR94}%
  \BibitemOpen
  \bibfield  {author} {\bibinfo {author} {\bibfnamefont {B.~H.}\ \bibnamefont
  {Greeley}}, \bibinfo {author} {\bibfnamefont {T.~V.}\ \bibnamefont {Russo}},
  \bibinfo {author} {\bibfnamefont {D.~T.}\ \bibnamefont {Mainz}}, \bibinfo
  {author} {\bibfnamefont {R.~A.}\ \bibnamefont {Friesner}}, \bibinfo {author}
  {\bibfnamefont {J.}~\bibnamefont {Langlois}}, \bibinfo {author}
  {\bibfnamefont {W.~A.}\ \bibnamefont {Goddard}, \bibfnamefont {III}},
  \bibinfo {author} {\bibfnamefont {R.~E.}\ \bibnamefont {Donnelly},
  \bibfnamefont {Jr.}}, \ and\ \bibinfo {author} {\bibfnamefont {M.~N.}\
  \bibnamefont {Ringnalda}},\ }\href {\doibase 10.1063/1.467520} {\bibfield
  {journal} {\bibinfo  {journal} {J. Chem. Phys.}\ }\textbf {\bibinfo {volume}
  {101}},\ \bibinfo {pages} {4028} (\bibinfo {year} {1994})}\BibitemShut
  {NoStop}%
\bibitem [{\citenamefont {Termath}\ and\ \citenamefont {Handy}(1994)}]{TH94}%
  \BibitemOpen
  \bibfield  {author} {\bibinfo {author} {\bibfnamefont {V.}~\bibnamefont
  {Termath}}\ and\ \bibinfo {author} {\bibfnamefont {N.~C.}\ \bibnamefont
  {Handy}},\ }\href {\doibase 10.1016/0009-2614(94)01160-5} {\bibfield
  {journal} {\bibinfo  {journal} {Chem. Phys. Lett.}\ }\textbf {\bibinfo
  {volume} {230}},\ \bibinfo {pages} {17} (\bibinfo {year} {1994})}\BibitemShut
  {NoStop}%
\bibitem [{\citenamefont {Murphy}, \citenamefont {Pollard},\ and\ \citenamefont
  {Friesner}(1997)}]{MPF97}%
  \BibitemOpen
  \bibfield  {author} {\bibinfo {author} {\bibfnamefont {R.~B.}\ \bibnamefont
  {Murphy}}, \bibinfo {author} {\bibfnamefont {W.~T.}\ \bibnamefont {Pollard}},
  \ and\ \bibinfo {author} {\bibfnamefont {R.~A.}\ \bibnamefont {Friesner}},\
  }\href {\doibase 10.1063/1.473553} {\bibfield  {journal} {\bibinfo  {journal}
  {J. Chem. Phys.}\ }\textbf {\bibinfo {volume} {106}},\ \bibinfo {pages}
  {5073} (\bibinfo {year} {1997})}\BibitemShut {NoStop}%
\bibitem [{\citenamefont {Izs\'ak}\ and\ \citenamefont {Neese}(2011)}]{IN11}%
  \BibitemOpen
  \bibfield  {author} {\bibinfo {author} {\bibfnamefont {R.}~\bibnamefont
  {Izs\'ak}}\ and\ \bibinfo {author} {\bibfnamefont {F.}~\bibnamefont
  {Neese}},\ }\href {\doibase 10.1063/1.3646921} {\bibfield  {journal}
  {\bibinfo  {journal} {J. Chem. Phys.}\ }\textbf {\bibinfo {volume} {135}},\
  \bibinfo {pages} {144105} (\bibinfo {year} {2011})}\BibitemShut {NoStop}%
\bibitem [{\citenamefont {Martinez}\ and\ \citenamefont {Carter}(1993)}]{MC93}%
  \BibitemOpen
  \bibfield  {author} {\bibinfo {author} {\bibfnamefont {T.~J.}\ \bibnamefont
  {Martinez}}\ and\ \bibinfo {author} {\bibfnamefont {E.~A.}\ \bibnamefont
  {Carter}},\ }\href {\doibase 10.1063/1.464751} {\bibfield  {journal}
  {\bibinfo  {journal} {J. Chem. Phys.}\ }\textbf {\bibinfo {volume} {98}},\
  \bibinfo {pages} {7081} (\bibinfo {year} {1993})}\BibitemShut {NoStop}%
\bibitem [{\citenamefont {Schrader}\ and\ \citenamefont {Prager}(1963)}]{SP63}%
  \BibitemOpen
  \bibfield  {author} {\bibinfo {author} {\bibfnamefont {D.~M.}\ \bibnamefont
  {Schrader}}\ and\ \bibinfo {author} {\bibfnamefont {S.}~\bibnamefont
  {Prager}},\ }\href {\doibase 10.1063/1.1733305} {\bibfield  {journal}
  {\bibinfo  {journal} {J. Chem. Phys.}\ }\textbf {\bibinfo {volume} {37}},\
  \bibinfo {pages} {1456} (\bibinfo {year} {1963})}\BibitemShut {NoStop}%
\bibitem [{\citenamefont {Whitten}(1973)}]{W73}%
  \BibitemOpen
  \bibfield  {author} {\bibinfo {author} {\bibfnamefont {J.~L.}\ \bibnamefont
  {Whitten}},\ }\href {\doibase 10.1063/1.1679012} {\bibfield  {journal}
  {\bibinfo  {journal} {J. Chem. Phys.}\ }\textbf {\bibinfo {volume} {58}},\
  \bibinfo {pages} {4496} (\bibinfo {year} {1973})}\BibitemShut {NoStop}%
\bibitem [{\citenamefont {Beebe}\ and\ \citenamefont
  {Linderberg}(1977)}]{BL77}%
  \BibitemOpen
  \bibfield  {author} {\bibinfo {author} {\bibfnamefont {N.~H.~F.}\
  \bibnamefont {Beebe}}\ and\ \bibinfo {author} {\bibfnamefont
  {J.}~\bibnamefont {Linderberg}},\ }\href {\doibase 10.1002/qua.560120408}
  {\bibfield  {journal} {\bibinfo  {journal} {Int. J. Quantum Chem.}\ }\textbf
  {\bibinfo {volume} {12}},\ \bibinfo {pages} {683} (\bibinfo {year}
  {1977})}\BibitemShut {NoStop}%
\bibitem [{\citenamefont {Alsenoy}(1988)}]{A88}%
  \BibitemOpen
  \bibfield  {author} {\bibinfo {author} {\bibfnamefont {C.~V.}\ \bibnamefont
  {Alsenoy}},\ }\href {\doibase 10.1002/jcc.540090607} {\bibfield  {journal}
  {\bibinfo  {journal} {J. Comp. Chem.}\ }\textbf {\bibinfo {volume} {9}},\
  \bibinfo {pages} {620} (\bibinfo {year} {1988})}\BibitemShut {NoStop}%
\bibitem [{\citenamefont {Vahtras}, \citenamefont {Alml\"of},\ and\
  \citenamefont {Feyereisen}(1993)}]{VAF93}%
  \BibitemOpen
  \bibfield  {author} {\bibinfo {author} {\bibfnamefont {O.}~\bibnamefont
  {Vahtras}}, \bibinfo {author} {\bibfnamefont {J.}~\bibnamefont {Alml\"of}}, \
  and\ \bibinfo {author} {\bibfnamefont {M.~W.}\ \bibnamefont {Feyereisen}},\
  }\href {\doibase 10.1016/0009-2614(93)89151-7} {\bibfield  {journal}
  {\bibinfo  {journal} {Chem. Phys. Lett.}\ }\textbf {\bibinfo {volume}
  {213}},\ \bibinfo {pages} {514} (\bibinfo {year} {1993})}\BibitemShut
  {NoStop}%
\bibitem [{\citenamefont {Feyereisen}, \citenamefont {Fitzgerald},\ and\
  \citenamefont {Komornicki}(1993)}]{FFK93}%
  \BibitemOpen
  \bibfield  {author} {\bibinfo {author} {\bibfnamefont {M.}~\bibnamefont
  {Feyereisen}}, \bibinfo {author} {\bibfnamefont {G.}~\bibnamefont
  {Fitzgerald}}, \ and\ \bibinfo {author} {\bibfnamefont {A.}~\bibnamefont
  {Komornicki}},\ }\href {\doibase 10.1016/0009-2614(93)87156-W} {\bibfield
  {journal} {\bibinfo  {journal} {Chem. Phys. Lett.}\ }\textbf {\bibinfo
  {volume} {208}},\ \bibinfo {pages} {359} (\bibinfo {year}
  {1993})}\BibitemShut {NoStop}%
\bibitem [{\citenamefont {Weigend}\ and\ \citenamefont {H\"aser}(1997)}]{WH97}%
  \BibitemOpen
  \bibfield  {author} {\bibinfo {author} {\bibfnamefont {F.}~\bibnamefont
  {Weigend}}\ and\ \bibinfo {author} {\bibfnamefont {M.}~\bibnamefont
  {H\"aser}},\ }\href {\doibase 10.1007/s002140050269} {\bibfield  {journal}
  {\bibinfo  {journal} {Theor. Chem. Acc.}\ }\textbf {\bibinfo {volume} {97}},\
  \bibinfo {pages} {331} (\bibinfo {year} {1997})}\BibitemShut {NoStop}%
\bibitem [{\citenamefont {Alml\"of}\ and\ \citenamefont {Taylor}(1987)}]{AT87}%
  \BibitemOpen
  \bibfield  {author} {\bibinfo {author} {\bibfnamefont {J.}~\bibnamefont
  {Alml\"of}}\ and\ \bibinfo {author} {\bibfnamefont {P.~R.}\ \bibnamefont
  {Taylor}},\ }\href {\doibase 10.1063/1.451917} {\bibfield  {journal}
  {\bibinfo  {journal} {J. Chem. Phys.}\ }\textbf {\bibinfo {volume} {86}},\
  \bibinfo {pages} {4070} (\bibinfo {year} {1987})}\BibitemShut {NoStop}%
\bibitem [{\citenamefont {Raffenetti}(1973)}]{R73}%
  \BibitemOpen
  \bibfield  {author} {\bibinfo {author} {\bibfnamefont {R.~C.}\ \bibnamefont
  {Raffenetti}},\ }\href {\doibase 10.1063/1.1679007} {\bibfield  {journal}
  {\bibinfo  {journal} {J. Chem. Phys.}\ }\textbf {\bibinfo {volume} {58}},\
  \bibinfo {pages} {4452} (\bibinfo {year} {1973})}\BibitemShut {NoStop}%
\bibitem [{\citenamefont {Alml\"of}\ and\ \citenamefont {Taylor}(1990)}]{AT90}%
  \BibitemOpen
  \bibfield  {author} {\bibinfo {author} {\bibfnamefont {J.}~\bibnamefont
  {Alml\"of}}\ and\ \bibinfo {author} {\bibfnamefont {P.~R.}\ \bibnamefont
  {Taylor}},\ }\href {\doibase 10.1063/1.458458} {\bibfield  {journal}
  {\bibinfo  {journal} {J. Chem. Phys.}\ }\textbf {\bibinfo {volume} {92}},\
  \bibinfo {pages} {551} (\bibinfo {year} {1990})}\BibitemShut {NoStop}%
\bibitem [{\citenamefont {Suaud}\ and\ \citenamefont {Malrieu}(2017)}]{SM17}%
  \BibitemOpen
  \bibfield  {author} {\bibinfo {author} {\bibfnamefont {N.}~\bibnamefont
  {Suaud}}\ and\ \bibinfo {author} {\bibfnamefont {J.~P.}\ \bibnamefont
  {Malrieu}},\ }\href {\doibase 10.1080/00268976.2017.1303207} {\bibfield
  {journal} {\bibinfo  {journal} {Mol. Phys.}\ }\textbf {\bibinfo {volume}
  {115}},\ \bibinfo {pages} {2684} (\bibinfo {year} {2017})}\BibitemShut
  {NoStop}%
\bibitem [{\citenamefont {Dunning}(1989)}]{D89}%
  \BibitemOpen
  \bibfield  {author} {\bibinfo {author} {\bibfnamefont {T.~H.}\ \bibnamefont
  {Dunning}},\ }\href {\doibase 10.1063/1.456153} {\bibfield  {journal}
  {\bibinfo  {journal} {J. Chem. Phys.}\ }\textbf {\bibinfo {volume} {90}},\
  \bibinfo {pages} {1007} (\bibinfo {year} {1989})}\BibitemShut {NoStop}%
\bibitem [{\citenamefont {Woon}\ and\ \citenamefont {Dunning}(1993)}]{WD93}%
  \BibitemOpen
  \bibfield  {author} {\bibinfo {author} {\bibfnamefont {D.~E.}\ \bibnamefont
  {Woon}}\ and\ \bibinfo {author} {\bibfnamefont {T.~H.}\ \bibnamefont
  {Dunning}},\ }\href {\doibase 10.1063/1.464303} {\bibfield  {journal}
  {\bibinfo  {journal} {J. Chem. Phys.}\ }\textbf {\bibinfo {volume} {98}},\
  \bibinfo {pages} {1358} (\bibinfo {year} {1993})}\BibitemShut {NoStop}%
\bibitem [{\citenamefont {Woon}\ and\ \citenamefont {Dunning}(1995)}]{WD95}%
  \BibitemOpen
  \bibfield  {author} {\bibinfo {author} {\bibfnamefont {D.~E.}\ \bibnamefont
  {Woon}}\ and\ \bibinfo {author} {\bibfnamefont {T.~H.}\ \bibnamefont
  {Dunning}},\ }\href {\doibase 10.1063/1.470645} {\bibfield  {journal}
  {\bibinfo  {journal} {J. Chem. Phys.}\ }\textbf {\bibinfo {volume} {103}},\
  \bibinfo {pages} {4572} (\bibinfo {year} {1995})}\BibitemShut {NoStop}%
\bibitem [{\citenamefont {Peterson}\ and\ \citenamefont
  {Dunning}(2002)}]{PD02}%
  \BibitemOpen
  \bibfield  {author} {\bibinfo {author} {\bibfnamefont {K.~A.}\ \bibnamefont
  {Peterson}}\ and\ \bibinfo {author} {\bibfnamefont {T.~H.}\ \bibnamefont
  {Dunning}},\ }\href {\doibase 10.1063/1.1520138} {\bibfield  {journal}
  {\bibinfo  {journal} {J. Chem. Phys.}\ }\textbf {\bibinfo {volume} {117}},\
  \bibinfo {pages} {10548} (\bibinfo {year} {2002})}\BibitemShut {NoStop}%
\bibitem [{\citenamefont {Lakin}(1965)}]{L65}%
  \BibitemOpen
  \bibfield  {author} {\bibinfo {author} {\bibfnamefont {W.}~\bibnamefont
  {Lakin}},\ }\href {\doibase 10.1063/1.1697255} {\bibfield  {journal}
  {\bibinfo  {journal} {J. Chem. Phys.}\ }\textbf {\bibinfo {volume} {43}},\
  \bibinfo {pages} {2954} (\bibinfo {year} {1965})}\BibitemShut {NoStop}%
\bibitem [{\citenamefont {Hill}(1985)}]{H85}%
  \BibitemOpen
  \bibfield  {author} {\bibinfo {author} {\bibfnamefont {R.~N.}\ \bibnamefont
  {Hill}},\ }\href {\doibase 10.1063/1.449481} {\bibfield  {journal} {\bibinfo
  {journal} {J. Chem. Phys.}\ }\textbf {\bibinfo {volume} {83}},\ \bibinfo
  {pages} {1173} (\bibinfo {year} {1985})}\BibitemShut {NoStop}%
\bibitem [{\citenamefont {Martin}\ and\ \citenamefont {Taylor}(1997)}]{MT97}%
  \BibitemOpen
  \bibfield  {author} {\bibinfo {author} {\bibfnamefont {J.~M.~L.}\
  \bibnamefont {Martin}}\ and\ \bibinfo {author} {\bibfnamefont {P.~R.}\
  \bibnamefont {Taylor}},\ }\href {\doibase 10.1063/1.473918} {\bibfield
  {journal} {\bibinfo  {journal} {J. Chem. Phys.}\ }\textbf {\bibinfo {volume}
  {106}},\ \bibinfo {pages} {8620} (\bibinfo {year} {1997})}\BibitemShut
  {NoStop}%
\bibitem [{\citenamefont {Halkier}\ \emph {et~al.}(1998)\citenamefont
  {Halkier}, \citenamefont {Helgaker}, \citenamefont {J{\o}rgensen},
  \citenamefont {Klopper}, \citenamefont {Koch}, \citenamefont {Olsen},\ and\
  \citenamefont {Wilson}}]{HHJKKOW98}%
  \BibitemOpen
  \bibfield  {author} {\bibinfo {author} {\bibfnamefont {A.}~\bibnamefont
  {Halkier}}, \bibinfo {author} {\bibfnamefont {T.}~\bibnamefont {Helgaker}},
  \bibinfo {author} {\bibfnamefont {P.}~\bibnamefont {J{\o}rgensen}}, \bibinfo
  {author} {\bibfnamefont {W.}~\bibnamefont {Klopper}}, \bibinfo {author}
  {\bibfnamefont {H.}~\bibnamefont {Koch}}, \bibinfo {author} {\bibfnamefont
  {J.}~\bibnamefont {Olsen}}, \ and\ \bibinfo {author} {\bibfnamefont {A.~K.}\
  \bibnamefont {Wilson}},\ }\href {\doibase 10.1016/S0009-2614(98)00111-0}
  {\bibfield  {journal} {\bibinfo  {journal} {Chem. Phys. Lett.}\ }\textbf
  {\bibinfo {volume} {286}},\ \bibinfo {pages} {243} (\bibinfo {year}
  {1998})}\BibitemShut {NoStop}%
\bibitem [{\citenamefont {Varandas}(2007)}]{V07}%
  \BibitemOpen
  \bibfield  {author} {\bibinfo {author} {\bibfnamefont {A.~J.~C.}\
  \bibnamefont {Varandas}},\ }\href {\doibase 10.1063/1.2741259} {\bibfield
  {journal} {\bibinfo  {journal} {J. Chem. Phys.}\ }\textbf {\bibinfo {volume}
  {126}},\ \bibinfo {pages} {244105} (\bibinfo {year} {2007})}\BibitemShut
  {NoStop}%
\bibitem [{\citenamefont {Hill}\ \emph {et~al.}(2009)\citenamefont {Hill},
  \citenamefont {Peterson}, \citenamefont {Knizia},\ and\ \citenamefont
  {Werner}}]{HPKW09}%
  \BibitemOpen
  \bibfield  {author} {\bibinfo {author} {\bibfnamefont {J.~G.}\ \bibnamefont
  {Hill}}, \bibinfo {author} {\bibfnamefont {K.~A.}\ \bibnamefont {Peterson}},
  \bibinfo {author} {\bibfnamefont {G.}~\bibnamefont {Knizia}}, \ and\ \bibinfo
  {author} {\bibfnamefont {H.-J.}\ \bibnamefont {Werner}},\ }\href {\doibase
  10.1063/1.3265857} {\bibfield  {journal} {\bibinfo  {journal} {J. Chem.
  Phys.}\ }\textbf {\bibinfo {volume} {131}},\ \bibinfo {pages} {194105}
  (\bibinfo {year} {2009})}\BibitemShut {NoStop}%
\bibitem [{\citenamefont {Kendall}, \citenamefont {Dunning},\ and\
  \citenamefont {Harrison}(1992)}]{KDH92}%
  \BibitemOpen
  \bibfield  {author} {\bibinfo {author} {\bibfnamefont {R.~A.}\ \bibnamefont
  {Kendall}}, \bibinfo {author} {\bibfnamefont {T.~H.}\ \bibnamefont
  {Dunning}}, \ and\ \bibinfo {author} {\bibfnamefont {R.~J.}\ \bibnamefont
  {Harrison}},\ }\href {\doibase 10.1063/1.462569} {\bibfield  {journal}
  {\bibinfo  {journal} {J. Chem. Phys.}\ }\textbf {\bibinfo {volume} {96}},\
  \bibinfo {pages} {6796} (\bibinfo {year} {1992})}\BibitemShut {NoStop}%
\bibitem [{\citenamefont {Woon}\ and\ \citenamefont {Dunning}(1994)}]{WD94}%
  \BibitemOpen
  \bibfield  {author} {\bibinfo {author} {\bibfnamefont {D.~E.}\ \bibnamefont
  {Woon}}\ and\ \bibinfo {author} {\bibfnamefont {T.~H.}\ \bibnamefont
  {Dunning}},\ }\href {\doibase 10.1063/1.466439} {\bibfield  {journal}
  {\bibinfo  {journal} {J. Chem. Phys.}\ }\textbf {\bibinfo {volume} {100}},\
  \bibinfo {pages} {2975} (\bibinfo {year} {1994})}\BibitemShut {NoStop}%
\bibitem [{\citenamefont {T.~H.~Dunning}(2001)}]{DPW01}%
  \BibitemOpen
  \bibfield  {author} {\bibinfo {author} {\bibfnamefont {A.~K.~W.}\
  \bibnamefont {T.~H.~Dunning}, \bibfnamefont {K.~A.~Peterson}},\ }\href
  {\doibase 10.1063/1.1367373} {\bibfield  {journal} {\bibinfo  {journal} {J.
  Chem. Phys.}\ }\textbf {\bibinfo {volume} {114}},\ \bibinfo {pages} {9244}
  (\bibinfo {year} {2001})}\BibitemShut {NoStop}%
\bibitem [{\citenamefont {Hashimoto}, \citenamefont {Hirao},\ and\
  \citenamefont {Tatewaki}(1997)}]{HHT97}%
  \BibitemOpen
  \bibfield  {author} {\bibinfo {author} {\bibfnamefont {T.}~\bibnamefont
  {Hashimoto}}, \bibinfo {author} {\bibfnamefont {K.}~\bibnamefont {Hirao}}, \
  and\ \bibinfo {author} {\bibfnamefont {H.}~\bibnamefont {Tatewaki}},\ }\href
  {\doibase 10.1016/S0009-2614(97)00613-1} {\bibfield  {journal} {\bibinfo
  {journal} {Chem. Phys. Lett.}\ }\textbf {\bibinfo {volume} {273}},\ \bibinfo
  {pages} {345} (\bibinfo {year} {1997})}\BibitemShut {NoStop}%
\bibitem [{\citenamefont {Chong}(1995)}]{C95}%
  \BibitemOpen
  \bibfield  {author} {\bibinfo {author} {\bibfnamefont {D.~P.}\ \bibnamefont
  {Chong}},\ }\href@noop {} {\bibfield  {journal} {\bibinfo  {journal} {Can. J.
  Chem.}\ }\textbf {\bibinfo {volume} {73}},\ \bibinfo {pages} {79} (\bibinfo
  {year} {1995})}\BibitemShut {NoStop}%
\bibitem [{\citenamefont {Manninen}\ and\ \citenamefont {Vaara}(2006)}]{MV06}%
  \BibitemOpen
  \bibfield  {author} {\bibinfo {author} {\bibfnamefont {P.}~\bibnamefont
  {Manninen}}\ and\ \bibinfo {author} {\bibfnamefont {J.}~\bibnamefont
  {Vaara}},\ }\href {\doibase 10.1002/jcc.20358} {\bibfield  {journal}
  {\bibinfo  {journal} {J. Comp. Chem.}\ }\textbf {\bibinfo {volume} {27}},\
  \bibinfo {pages} {434} (\bibinfo {year} {2006})}\BibitemShut {NoStop}%
\bibitem [{\citenamefont {Lehtola}\ \emph {et~al.}(2013)\citenamefont
  {Lehtola}, \citenamefont {Manninen}, \citenamefont {Hakala},\ and\
  \citenamefont {Hamalainen}}]{LMHH13}%
  \BibitemOpen
  \bibfield  {author} {\bibinfo {author} {\bibfnamefont {S.}~\bibnamefont
  {Lehtola}}, \bibinfo {author} {\bibfnamefont {P.}~\bibnamefont {Manninen}},
  \bibinfo {author} {\bibfnamefont {M.}~\bibnamefont {Hakala}}, \ and\ \bibinfo
  {author} {\bibfnamefont {K.}~\bibnamefont {Hamalainen}},\ }\href {\doibase
  10.1063/1.4788635} {\bibfield  {journal} {\bibinfo  {journal} {J. Chem.
  Phys.}\ }\textbf {\bibinfo {volume} {138}},\ \bibinfo {pages} {044109}
  (\bibinfo {year} {2013})}\BibitemShut {NoStop}%
\bibitem [{\citenamefont {Delley}(1990)}]{D90}%
  \BibitemOpen
  \bibfield  {author} {\bibinfo {author} {\bibfnamefont {B.}~\bibnamefont
  {Delley}},\ }\href {\doibase 10.1063/1.458452} {\bibfield  {journal}
  {\bibinfo  {journal} {J. Chem. Phys.}\ }\textbf {\bibinfo {volume} {92}},\
  \bibinfo {pages} {508} (\bibinfo {year} {1990})}\BibitemShut {NoStop}%
\bibitem [{\citenamefont {Jensen}(2001)}]{J01}%
  \BibitemOpen
  \bibfield  {author} {\bibinfo {author} {\bibfnamefont {F.}~\bibnamefont
  {Jensen}},\ }\href {\doibase 10.1063/1.1413524} {\bibfield  {journal}
  {\bibinfo  {journal} {J. Chem. Phys.}\ }\textbf {\bibinfo {volume} {115}},\
  \bibinfo {pages} {9113} (\bibinfo {year} {2001})}\BibitemShut {NoStop}%
\bibitem [{\citenamefont {Jensen}(2002{\natexlab{a}})}]{J02b}%
  \BibitemOpen
  \bibfield  {author} {\bibinfo {author} {\bibfnamefont {F.}~\bibnamefont
  {Jensen}},\ }\href {\doibase 10.1063/1.1515484} {\bibfield  {journal}
  {\bibinfo  {journal} {J. Chem. Phys.}\ }\textbf {\bibinfo {volume} {117}},\
  \bibinfo {pages} {9234} (\bibinfo {year} {2002}{\natexlab{a}})}\BibitemShut
  {NoStop}%
\bibitem [{\citenamefont {Jensen}(2002{\natexlab{b}})}]{J02a}%
  \BibitemOpen
  \bibfield  {author} {\bibinfo {author} {\bibfnamefont {F.}~\bibnamefont
  {Jensen}},\ }\href {\doibase 10.1063/1.1465405} {\bibfield  {journal}
  {\bibinfo  {journal} {J. Chem. Phys.}\ }\textbf {\bibinfo {volume} {116}},\
  \bibinfo {pages} {7372} (\bibinfo {year} {2002}{\natexlab{b}})}\BibitemShut
  {NoStop}%
\bibitem [{\citenamefont {VandeVondele}\ and\ \citenamefont
  {Hutter}(2007)}]{VH07}%
  \BibitemOpen
  \bibfield  {author} {\bibinfo {author} {\bibfnamefont {J.}~\bibnamefont
  {VandeVondele}}\ and\ \bibinfo {author} {\bibfnamefont {J.}~\bibnamefont
  {Hutter}},\ }\href {\doibase 10.1063/1.2770708} {\bibfield  {journal}
  {\bibinfo  {journal} {J. Chem. Phys.}\ }\textbf {\bibinfo {volume} {127}},\
  \bibinfo {pages} {114105} (\bibinfo {year} {2007})}\BibitemShut {NoStop}%
\bibitem [{\citenamefont {van Duijneveldt-van~de Rijdt}\ and\ \citenamefont
  {van Duijneveldt}(1999)}]{DD99}%
  \BibitemOpen
  \bibfield  {author} {\bibinfo {author} {\bibfnamefont {J.~G. C.~M.}\
  \bibnamefont {van Duijneveldt-van~de Rijdt}}\ and\ \bibinfo {author}
  {\bibfnamefont {F.~B.}\ \bibnamefont {van Duijneveldt}},\ }\href {\doibase
  10.1063/1.479684} {\bibfield  {journal} {\bibinfo  {journal} {J. Chem.
  Phys.}\ }\textbf {\bibinfo {volume} {111}},\ \bibinfo {pages} {3812}
  (\bibinfo {year} {1999})}\BibitemShut {NoStop}%
\bibitem [{\citenamefont {Boys}\ and\ \citenamefont {Bernardi}(1970)}]{BB70}%
  \BibitemOpen
  \bibfield  {author} {\bibinfo {author} {\bibfnamefont {S.~F.}\ \bibnamefont
  {Boys}}\ and\ \bibinfo {author} {\bibfnamefont {F.}~\bibnamefont
  {Bernardi}},\ }\href {\doibase 10.1080/00268977000101561} {\bibfield
  {journal} {\bibinfo  {journal} {Mol. Phys.}\ }\textbf {\bibinfo {volume}
  {19}},\ \bibinfo {pages} {553} (\bibinfo {year} {1970})}\BibitemShut
  {NoStop}%
\bibitem [{\citenamefont {Laikov}(2005)}]{L05}%
  \BibitemOpen
  \bibfield  {author} {\bibinfo {author} {\bibfnamefont {D.~N.}\ \bibnamefont
  {Laikov}},\ }\href {\doibase 10.1016/j.cplett.2005.09.046} {\bibfield
  {journal} {\bibinfo  {journal} {Chem. Phys. Lett.}\ }\textbf {\bibinfo
  {volume} {416}},\ \bibinfo {pages} {116} (\bibinfo {year}
  {2005})}\BibitemShut {NoStop}%
\bibitem [{\citenamefont {Adamowicz}\ and\ \citenamefont
  {Bartlett}(1987)}]{AB87}%
  \BibitemOpen
  \bibfield  {author} {\bibinfo {author} {\bibfnamefont {L.}~\bibnamefont
  {Adamowicz}}\ and\ \bibinfo {author} {\bibfnamefont {R.~J.}\ \bibnamefont
  {Bartlett}},\ }\href {\doibase 10.1063/1.452468} {\bibfield  {journal}
  {\bibinfo  {journal} {J. Chem. Phys.}\ }\textbf {\bibinfo {volume} {86}},\
  \bibinfo {pages} {6314} (\bibinfo {year} {1987})}\BibitemShut {NoStop}%
\bibitem [{\citenamefont {Dyall}(1994)}]{D94}%
  \BibitemOpen
  \bibfield  {author} {\bibinfo {author} {\bibfnamefont {K.~G.}\ \bibnamefont
  {Dyall}},\ }\href {\doibase 10.1063/1.466508} {\bibfield  {journal} {\bibinfo
   {journal} {J. Chem. Phys.}\ }\textbf {\bibinfo {volume} {100}},\ \bibinfo
  {pages} {2118} (\bibinfo {year} {1994})}\BibitemShut {NoStop}%
\bibitem [{\citenamefont {Shamov}, \citenamefont {Schreckenbach},\ and\
  \citenamefont {Vo}(2007)}]{SSV07}%
  \BibitemOpen
  \bibfield  {author} {\bibinfo {author} {\bibfnamefont {G.~A.}\ \bibnamefont
  {Shamov}}, \bibinfo {author} {\bibfnamefont {G.}~\bibnamefont
  {Schreckenbach}}, \ and\ \bibinfo {author} {\bibfnamefont {T.~N.}\
  \bibnamefont {Vo}},\ }\href {\doibase 10.1002/chem.200601244} {\bibfield
  {journal} {\bibinfo  {journal} {Chem. Eur. J.}\ }\textbf {\bibinfo {volume}
  {13}},\ \bibinfo {pages} {4932} (\bibinfo {year} {2007})}\BibitemShut
  {NoStop}%
\bibitem [{\citenamefont {Shamov}\ and\ \citenamefont
  {Schreckenbach}(2006)}]{SS06}%
  \BibitemOpen
  \bibfield  {author} {\bibinfo {author} {\bibfnamefont {G.~A.}\ \bibnamefont
  {Shamov}}\ and\ \bibinfo {author} {\bibfnamefont {G.}~\bibnamefont
  {Schreckenbach}},\ }\href {\doibase 10.1021/jp063060l} {\bibfield  {journal}
  {\bibinfo  {journal} {J. Phys. Chem. A}\ }\textbf {\bibinfo {volume} {110}},\
  \bibinfo {pages} {9486} (\bibinfo {year} {2006})}\BibitemShut {NoStop}%
\bibitem [{\citenamefont {Ustynyuk}\ \emph {et~al.}(2014)\citenamefont
  {Ustynyuk}, \citenamefont {Gloriozov}, \citenamefont {Kalmykov},
  \citenamefont {Mitrofanov}, \citenamefont {Babain}, \citenamefont
  {Alyapyshev},\ and\ \citenamefont {Ustynyuk}}]{UGKMBAU14}%
  \BibitemOpen
  \bibfield  {author} {\bibinfo {author} {\bibfnamefont {Y.~A.}\ \bibnamefont
  {Ustynyuk}}, \bibinfo {author} {\bibfnamefont {I.~P.}\ \bibnamefont
  {Gloriozov}}, \bibinfo {author} {\bibfnamefont {S.~N.}\ \bibnamefont
  {Kalmykov}}, \bibinfo {author} {\bibfnamefont {A.~A.}\ \bibnamefont
  {Mitrofanov}}, \bibinfo {author} {\bibfnamefont {V.~A.}\ \bibnamefont
  {Babain}}, \bibinfo {author} {\bibfnamefont {M.~Y.}\ \bibnamefont
  {Alyapyshev}}, \ and\ \bibinfo {author} {\bibfnamefont {N.~A.}\ \bibnamefont
  {Ustynyuk}},\ }\href {\doibase 10.1080/07366299.2014.915666} {\bibfield
  {journal} {\bibinfo  {journal} {Solv. Extr. Ion Exch.}\ }\textbf {\bibinfo
  {volume} {32}},\ \bibinfo {pages} {508} (\bibinfo {year} {2014})}\BibitemShut
  {NoStop}%
\bibitem [{\citenamefont {Ustynyuk}\ \emph {et~al.}(2017)\citenamefont
  {Ustynyuk}, \citenamefont {Kalmykov}, \citenamefont {Alyapyshev},
  \citenamefont {Babain}, \citenamefont {Lavrov}, \citenamefont {Zhokhov},
  \citenamefont {Matveev}, \citenamefont {Gloriozov}, \citenamefont
  {Tkachenko}, \citenamefont {Voronaev},\ and\ \citenamefont
  {Ustynyuk}}]{UKABLZMGTVU17}%
  \BibitemOpen
  \bibfield  {author} {\bibinfo {author} {\bibfnamefont {Y.~A.}\ \bibnamefont
  {Ustynyuk}}, \bibinfo {author} {\bibfnamefont {S.~N.}\ \bibnamefont
  {Kalmykov}}, \bibinfo {author} {\bibfnamefont {M.}~\bibnamefont
  {Alyapyshev}}, \bibinfo {author} {\bibfnamefont {V.}~\bibnamefont {Babain}},
  \bibinfo {author} {\bibfnamefont {H.~V.}\ \bibnamefont {Lavrov}}, \bibinfo
  {author} {\bibfnamefont {S.~S.}\ \bibnamefont {Zhokhov}}, \bibinfo {author}
  {\bibfnamefont {P.~I.}\ \bibnamefont {Matveev}}, \bibinfo {author}
  {\bibfnamefont {I.~P.}\ \bibnamefont {Gloriozov}}, \bibinfo {author}
  {\bibfnamefont {L.~I.}\ \bibnamefont {Tkachenko}}, \bibinfo {author}
  {\bibfnamefont {I.~G.}\ \bibnamefont {Voronaev}}, \ and\ \bibinfo {author}
  {\bibfnamefont {N.~A.}\ \bibnamefont {Ustynyuk}},\ }\href {\doibase
  10.1039/C7DT01009E} {\bibfield  {journal} {\bibinfo  {journal} {Dalton
  Trans.}\ }\textbf {\bibinfo {volume} {46}},\ \bibinfo {pages} {10926}
  (\bibinfo {year} {2017})}\BibitemShut {NoStop}%
\bibitem [{\citenamefont {Saenko}\ \emph {et~al.}(2011)\citenamefont {Saenko},
  \citenamefont {Laikov}, \citenamefont {Baranova},\ and\ \citenamefont
  {Feldman}}]{SLBF11}%
  \BibitemOpen
  \bibfield  {author} {\bibinfo {author} {\bibfnamefont {E.~V.}\ \bibnamefont
  {Saenko}}, \bibinfo {author} {\bibfnamefont {D.~N.}\ \bibnamefont {Laikov}},
  \bibinfo {author} {\bibfnamefont {I.~A.}\ \bibnamefont {Baranova}}, \ and\
  \bibinfo {author} {\bibfnamefont {V.~I.}\ \bibnamefont {Feldman}},\ }\href
  {\doibase 10.1063/1.3638690} {\bibfield  {journal} {\bibinfo  {journal} {J.
  Chem. Phys.}\ }\textbf {\bibinfo {volume} {135}},\ \bibinfo {pages} {101103}
  (\bibinfo {year} {2011})}\BibitemShut {NoStop}%
\bibitem [{\citenamefont {Handy}, \citenamefont {Marron},\ and\ \citenamefont
  {Silverstone}(1969)}]{HMS69}%
  \BibitemOpen
  \bibfield  {author} {\bibinfo {author} {\bibfnamefont {N.~C.}\ \bibnamefont
  {Handy}}, \bibinfo {author} {\bibfnamefont {M.~T.}\ \bibnamefont {Marron}}, \
  and\ \bibinfo {author} {\bibfnamefont {H.~J.}\ \bibnamefont {Silverstone}},\
  }\href {\doibase 10.1103/PhysRev.180.45} {\bibfield  {journal} {\bibinfo
  {journal} {Phys. Rev.}\ }\textbf {\bibinfo {volume} {180}},\ \bibinfo {pages}
  {45} (\bibinfo {year} {1969})}\BibitemShut {NoStop}%
\bibitem [{\citenamefont {Morrell}, \citenamefont {Parr},\ and\ \citenamefont
  {Levy}(1975)}]{MPL75}%
  \BibitemOpen
  \bibfield  {author} {\bibinfo {author} {\bibfnamefont {M.~M.}\ \bibnamefont
  {Morrell}}, \bibinfo {author} {\bibfnamefont {R.~G.}\ \bibnamefont {Parr}}, \
  and\ \bibinfo {author} {\bibfnamefont {M.}~\bibnamefont {Levy}},\ }\href
  {\doibase 10.1063/1.430509} {\bibfield  {journal} {\bibinfo  {journal} {J.
  Chem. Phys.}\ }\textbf {\bibinfo {volume} {62}},\ \bibinfo {pages} {549}
  (\bibinfo {year} {1975})}\BibitemShut {NoStop}%
\bibitem [{\citenamefont {Aoyama}\ \emph {et~al.}(2012)\citenamefont {Aoyama},
  \citenamefont {Hayakawa}, \citenamefont {Kinoshita},\ and\ \citenamefont
  {Nio}}]{AHKN12}%
  \BibitemOpen
  \bibfield  {author} {\bibinfo {author} {\bibfnamefont {T.}~\bibnamefont
  {Aoyama}}, \bibinfo {author} {\bibfnamefont {M.}~\bibnamefont {Hayakawa}},
  \bibinfo {author} {\bibfnamefont {T.}~\bibnamefont {Kinoshita}}, \ and\
  \bibinfo {author} {\bibfnamefont {M.}~\bibnamefont {Nio}},\ }\href {\doibase
  10.1103/PhysRevLett.109.111807} {\bibfield  {journal} {\bibinfo  {journal}
  {Phys. Rev. Lett.}\ }\textbf {\bibinfo {volume} {109}},\ \bibinfo {pages}
  {111807} (\bibinfo {year} {2012})}\BibitemShut {NoStop}%
\bibitem [{\citenamefont {Hanneke}, \citenamefont {Hoogerheide},\ and\
  \citenamefont {Gabrielse}(2011)}]{HFG11}%
  \BibitemOpen
  \bibfield  {author} {\bibinfo {author} {\bibfnamefont {D.}~\bibnamefont
  {Hanneke}}, \bibinfo {author} {\bibfnamefont {S.~F.}\ \bibnamefont
  {Hoogerheide}}, \ and\ \bibinfo {author} {\bibfnamefont {G.}~\bibnamefont
  {Gabrielse}},\ }\href {\doibase 10.1103/PhysRevA.83.052122} {\bibfield
  {journal} {\bibinfo  {journal} {Phys. Rev. A}\ }\textbf {\bibinfo {volume}
  {83}},\ \bibinfo {pages} {052122} (\bibinfo {year} {2011})}\BibitemShut
  {NoStop}%
\bibitem [{\citenamefont {Visscher}\ and\ \citenamefont {Dyall}(1997)}]{VD97}%
  \BibitemOpen
  \bibfield  {author} {\bibinfo {author} {\bibfnamefont {L.}~\bibnamefont
  {Visscher}}\ and\ \bibinfo {author} {\bibfnamefont {K.~G.}\ \bibnamefont
  {Dyall}},\ }\href {\doibase 10.1006/adnd.1997.0751} {\bibfield  {journal}
  {\bibinfo  {journal} {Atom. Data Nucl. Data}\ }\textbf {\bibinfo {volume}
  {67}},\ \bibinfo {pages} {207} (\bibinfo {year} {1997})}\BibitemShut
  {NoStop}%
\bibitem [{\citenamefont {Grant}, \citenamefont {Mayers},\ and\ \citenamefont
  {Pyper}(1976)}]{GMP76}%
  \BibitemOpen
  \bibfield  {author} {\bibinfo {author} {\bibfnamefont {I.~P.}\ \bibnamefont
  {Grant}}, \bibinfo {author} {\bibfnamefont {D.~F.}\ \bibnamefont {Mayers}}, \
  and\ \bibinfo {author} {\bibfnamefont {N.~C.}\ \bibnamefont {Pyper}},\ }\href
  {\doibase 10.1088/0022-3700/9/16/013} {\bibfield  {journal} {\bibinfo
  {journal} {J. Phys. B: At. Mol. Phys.}\ }\textbf {\bibinfo {volume} {9}},\
  \bibinfo {pages} {2777} (\bibinfo {year} {1976})}\BibitemShut {NoStop}%
\bibitem [{\citenamefont {Wesolowski}\ \emph {et~al.}(2000)\citenamefont
  {Wesolowski}, \citenamefont {Valeev}, \citenamefont {King}, \citenamefont
  {Baranovski},\ and\ \citenamefont {Schaefer}}]{WVKBS00}%
  \BibitemOpen
  \bibfield  {author} {\bibinfo {author} {\bibfnamefont {S.~S.}\ \bibnamefont
  {Wesolowski}}, \bibinfo {author} {\bibfnamefont {E.~F.}\ \bibnamefont
  {Valeev}}, \bibinfo {author} {\bibfnamefont {R.~A.}\ \bibnamefont {King}},
  \bibinfo {author} {\bibfnamefont {V.}~\bibnamefont {Baranovski}}, \ and\
  \bibinfo {author} {\bibfnamefont {H.~F.}\ \bibnamefont {Schaefer}},\ }\href
  {\doibase 10.1080/00268970050080582} {\bibfield  {journal} {\bibinfo
  {journal} {Mol. Phys.}\ }\textbf {\bibinfo {volume} {98}},\ \bibinfo {pages}
  {1227} (\bibinfo {year} {2000})}\BibitemShut {NoStop}%
\bibitem [{SM()}]{SM}%
  \BibitemOpen
  \href {http://jcp.air.org/} {}\bibinfo {note} {Supplementary
  material}\BibitemShut {NoStop}%
\bibitem [{\citenamefont {Sosulin}\ \emph {et~al.}(2017)\citenamefont
  {Sosulin}, \citenamefont {Shiryaeva}, \citenamefont {Tyurin},\ and\
  \citenamefont {Feldman}}]{SSTF17}%
  \BibitemOpen
  \bibfield  {author} {\bibinfo {author} {\bibfnamefont {I.~S.}\ \bibnamefont
  {Sosulin}}, \bibinfo {author} {\bibfnamefont {E.~S.}\ \bibnamefont
  {Shiryaeva}}, \bibinfo {author} {\bibfnamefont {D.~A.}\ \bibnamefont
  {Tyurin}}, \ and\ \bibinfo {author} {\bibfnamefont {V.~I.}\ \bibnamefont
  {Feldman}},\ }\href {\doibase 10.1063/1.4999772} {\bibfield  {journal}
  {\bibinfo  {journal} {J. Chem. Phys.}\ }\textbf {\bibinfo {volume} {147}},\
  \bibinfo {pages} {131102} (\bibinfo {year} {2017})}\BibitemShut {NoStop}%
\bibitem [{\citenamefont {Ryazantsev}, \citenamefont {Tyurin},\ and\
  \citenamefont {Feldman}(2017)}]{RTF17}%
  \BibitemOpen
  \bibfield  {author} {\bibinfo {author} {\bibfnamefont {S.~V.}\ \bibnamefont
  {Ryazantsev}}, \bibinfo {author} {\bibfnamefont {D.~A.}\ \bibnamefont
  {Tyurin}}, \ and\ \bibinfo {author} {\bibfnamefont {V.~I.}\ \bibnamefont
  {Feldman}},\ }\href {\doibase 10.1016/j.saa.2017.06.018} {\bibfield
  {journal} {\bibinfo  {journal} {Spectrochim. Acta A}\ }\textbf {\bibinfo
  {volume} {187}},\ \bibinfo {pages} {39} (\bibinfo {year} {2017})}\BibitemShut
  {NoStop}%
\bibitem [{\citenamefont {Ryazantsev}\ \emph {et~al.}(2017)\citenamefont
  {Ryazantsev}, \citenamefont {Tyurin}, \citenamefont {Feldman},\ and\
  \citenamefont {Khriachtchev}}]{RTFK17}%
  \BibitemOpen
  \bibfield  {author} {\bibinfo {author} {\bibfnamefont {S.~V.}\ \bibnamefont
  {Ryazantsev}}, \bibinfo {author} {\bibfnamefont {D.~A.}\ \bibnamefont
  {Tyurin}}, \bibinfo {author} {\bibfnamefont {V.~I.}\ \bibnamefont {Feldman}},
  \ and\ \bibinfo {author} {\bibfnamefont {L.}~\bibnamefont {Khriachtchev}},\
  }\href {\doibase 10.1063/1.5000578} {\bibfield  {journal} {\bibinfo
  {journal} {J. Chem. Phys.}\ }\textbf {\bibinfo {volume} {147}},\ \bibinfo
  {pages} {184301} (\bibinfo {year} {2017})}\BibitemShut {NoStop}%
\bibitem [{\citenamefont {Kameneva}, \citenamefont {Tyurin},\ and\
  \citenamefont {Feldman}(2017)}]{KTF17}%
  \BibitemOpen
  \bibfield  {author} {\bibinfo {author} {\bibfnamefont {S.~V.}\ \bibnamefont
  {Kameneva}}, \bibinfo {author} {\bibfnamefont {D.~A.}\ \bibnamefont
  {Tyurin}}, \ and\ \bibinfo {author} {\bibfnamefont {V.~I.}\ \bibnamefont
  {Feldman}},\ }\href {\doibase 10.1039/c7cp03518g} {\bibfield  {journal}
  {\bibinfo  {journal} {Phys. Chem. Chem. Phys.}\ }\textbf {\bibinfo {volume}
  {19}},\ \bibinfo {pages} {24348} (\bibinfo {year} {2017})}\BibitemShut
  {NoStop}%
\bibitem [{\citenamefont {Kameneva}\ \emph {et~al.}(2016)\citenamefont
  {Kameneva}, \citenamefont {Tyurin}, \citenamefont {Nuzhdin},\ and\
  \citenamefont {Feldman}}]{KTNF16}%
  \BibitemOpen
  \bibfield  {author} {\bibinfo {author} {\bibfnamefont {S.~V.}\ \bibnamefont
  {Kameneva}}, \bibinfo {author} {\bibfnamefont {D.~A.}\ \bibnamefont
  {Tyurin}}, \bibinfo {author} {\bibfnamefont {K.~B.}\ \bibnamefont {Nuzhdin}},
  \ and\ \bibinfo {author} {\bibfnamefont {V.~I.}\ \bibnamefont {Feldman}},\
  }\href {\doibase 10.1063/1.4969075} {\bibfield  {journal} {\bibinfo
  {journal} {J. Chem. Phys.}\ }\textbf {\bibinfo {volume} {145}},\ \bibinfo
  {pages} {214309} (\bibinfo {year} {2016})}\BibitemShut {NoStop}%
\bibitem [{\citenamefont {Shiryaeva}, \citenamefont {Tyurin},\ and\
  \citenamefont {Feldman}(2016)}]{STF16}%
  \BibitemOpen
  \bibfield  {author} {\bibinfo {author} {\bibfnamefont {E.~S.}\ \bibnamefont
  {Shiryaeva}}, \bibinfo {author} {\bibfnamefont {D.~A.}\ \bibnamefont
  {Tyurin}}, \ and\ \bibinfo {author} {\bibfnamefont {V.~I.}\ \bibnamefont
  {Feldman}},\ }\href {\doibase 10.1021/acs.jpca.6b07301} {\bibfield  {journal}
  {\bibinfo  {journal} {J. Phys. Chem. A}\ }\textbf {\bibinfo {volume} {120}},\
  \bibinfo {pages} {7847} (\bibinfo {year} {2016})}\BibitemShut {NoStop}%
\bibitem [{\citenamefont {Sosulin}\ \emph {et~al.}(2018)\citenamefont
  {Sosulin}, \citenamefont {Shiryaeva}, \citenamefont {Tyurin},\ and\
  \citenamefont {Feldman}}]{SSTF18}%
  \BibitemOpen
  \bibfield  {author} {\bibinfo {author} {\bibfnamefont {I.~S.}\ \bibnamefont
  {Sosulin}}, \bibinfo {author} {\bibfnamefont {E.~S.}\ \bibnamefont
  {Shiryaeva}}, \bibinfo {author} {\bibfnamefont {D.~A.}\ \bibnamefont
  {Tyurin}}, \ and\ \bibinfo {author} {\bibfnamefont {V.~I.}\ \bibnamefont
  {Feldman}},\ }\href {\doibase 10.1021/acs.jpca.8b01485} {\bibfield  {journal}
  {\bibinfo  {journal} {J. Phys. Chem. A}\ }\textbf {\bibinfo {volume} {122}},\
  \bibinfo {pages} {4042} (\bibinfo {year} {2018})}\BibitemShut {NoStop}%
\bibitem [{\citenamefont {Laikov}(2011)}]{L11}%
  \BibitemOpen
  \bibfield  {author} {\bibinfo {author} {\bibfnamefont {D.~N.}\ \bibnamefont
  {Laikov}},\ }\href {\doibase 10.1063/1.3646498} {\bibfield  {journal}
  {\bibinfo  {journal} {J. Chem. Phys.}\ }\textbf {\bibinfo {volume} {135}},\
  \bibinfo {pages} {134120} (\bibinfo {year} {2011})}\BibitemShut {NoStop}%
\bibitem [{\citenamefont {Briling}(2017)}]{B17}%
  \BibitemOpen
  \bibfield  {author} {\bibinfo {author} {\bibfnamefont {K.~R.}\ \bibnamefont
  {Briling}},\ }\href {\doibase 10.1063/1.5000525} {\bibfield  {journal}
  {\bibinfo  {journal} {J. Chem. Phys.}\ }\textbf {\bibinfo {volume} {147}},\
  \bibinfo {pages} {157101} (\bibinfo {year} {2017})}\BibitemShut {NoStop}%
\end{thebibliography}
\end{document}